\documentclass[a4paper, amsfonts, amssymb, amsmath, reprint, showkeys, nofootinbib, twoside,twocolumns,floatfix,superscriptaddress]{revtex4-1}
\usepackage[english]{babel}
\usepackage[utf8]{inputenc}
\usepackage{xcolor}
\usepackage[colorinlistoftodos, color=green!40, prependcaption]{todonotes}
\usepackage{todonotes}
\usepackage{amsthm}
\usepackage{mathtools}
\usepackage{physics}
\usepackage{xcolor}
\usepackage{graphicx}
\usepackage[left=13mm,right=13mm,top=35mm,columnsep=15pt]{geometry} 
\usepackage{adjustbox}
\usepackage{placeins}
\usepackage[T1]{fontenc}
\usepackage{lipsum}
\usepackage{csquotes}
\usepackage[pdftex, pdftitle={Article}, pdfauthor={Author}]{hyperref} % For hyperlinks in the PDF
\usepackage{amssymb}
\usepackage{mathtools}
\newcommand{\MMSb}{{\rm M}\overline{\mathrm{MS}}}
\newcommand{\NLO}{\rm NLO}

\begin{document}
\title{Parton distribution functions from lattice QCD using Bayes-Gauss-Fourier transforms}

\author{Constantia Alexandrou}
    %\email{email@institution.com}% Your name
    \affiliation{Department  of  Physics,  University  of  Cyprus,  P.O.  Box  20537,  1678  Nicosia,  Cyprus}
    \affiliation{The  Cyprus  Institute,  20  Konstantinou  Kavafi  Str.,  2121  Nicosia,  Cyprus}
    \author{Giovanni Iannelli}
    %\email{g.iannelli@stimulate-ejd.eu}
    \affiliation{Department  of  Physics,  University  of  Cyprus,  P.O.  Box  20537,  1678  Nicosia,  Cyprus}
    \affiliation{Università di Roma Tor Vergata, Via della Ricerca Scientifica 1, I-00133, Rome, Italy}
    \affiliation{Institut für  Physik, Humboldt-Universität zu Berlin, Newtonstr. 15, 12489 Berlin, Germany}
        \affiliation{Deutsches  Elektronen-Synchrotron,  15738  Zeuthen,  Germany}
    \author{Karl Jansen}
    \affiliation{Deutsches  Elektronen-Synchrotron,  15738  Zeuthen,  Germany}
    \author{Floriano Manigrasso}
    %\email{f.manigrasso@stimulate-ejd.eu}
    \affiliation{Department  of  Physics,  University  of  Cyprus,  P.O.  Box  20537,  1678  Nicosia,  Cyprus}
    \affiliation{Università di Roma Tor Vergata, Via della Ricerca Scientifica 1, I-00133, Rome, Italy}
    \affiliation{Institut für  Physik, Humboldt-Universität zu Berlin, Newtonstr. 15, 12489 Berlin, Germany
\centerline{\includegraphics[width=0.15\linewidth]{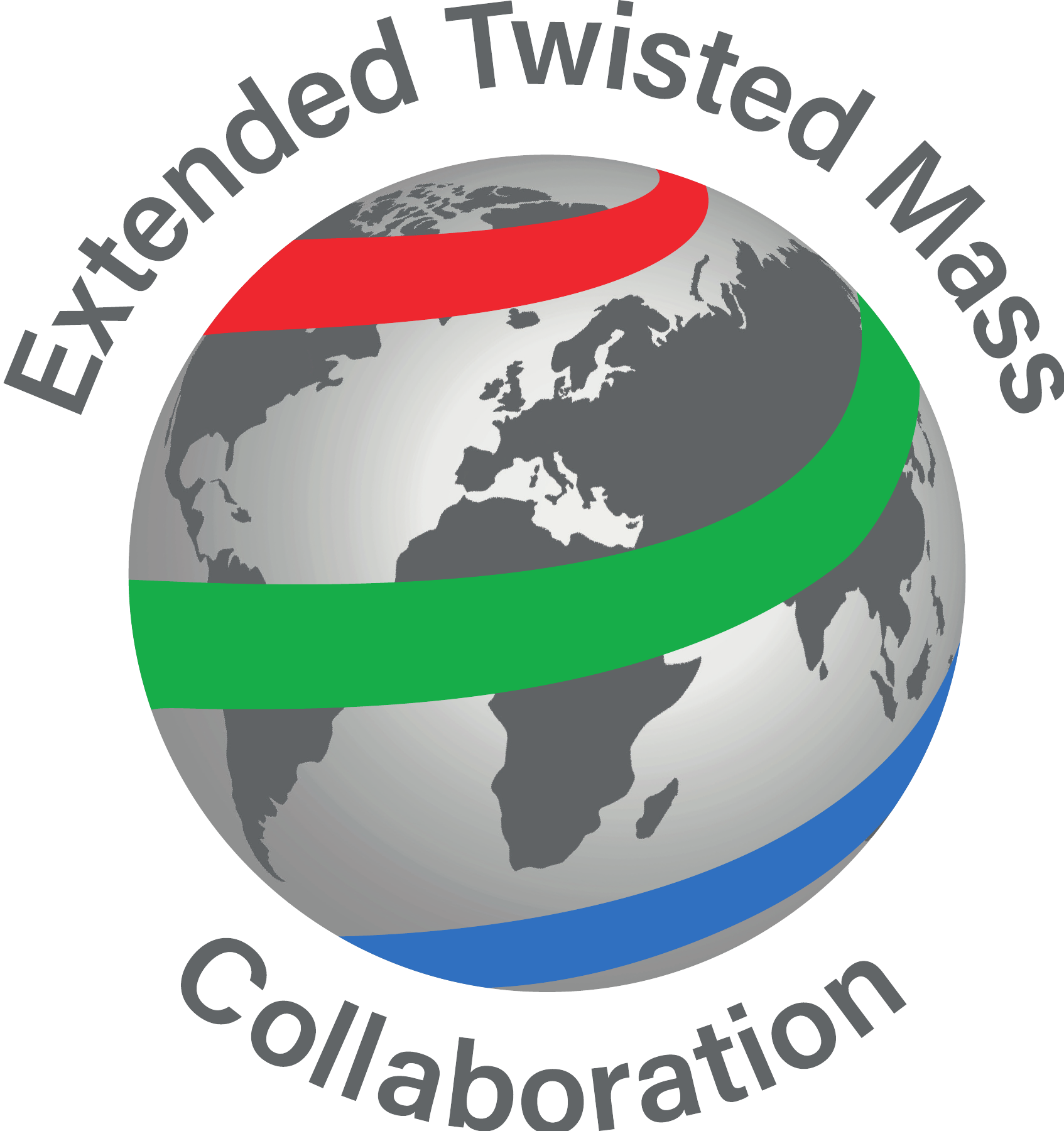}}
    }

\date{\today} % Leave empty to omit a date

\begin{abstract}
\vspace*{0.3cm}
 We present a new 
method, based on Gaussian process regression, for reconstructing the continuous 
$x$-dependence of parton distribution functions (PDFs) from quasi-PDFs computed using lattice QCD. 
We examine the origin of  the unphysical oscillations seen in current lattice calculations of quasi-PDFs and develop a nonparametric fitting approach to take the required Fourier transform.
The method is tested on one ensemble of maximally twisted mass fermions 
with two light quarks. 
We find that with our approach oscillations of the quasi-PDF are drastically 
reduced. However, the final effect on the light-cone PDFs is small. 
This finding suggests that the deviation seen between current lattice QCD results and phenomenological determinations cannot be attributed solely on the Fourier transform.
\end{abstract}


\maketitle

\section{Introduction} \label{sec:outline}

Parton distribution functions (PDFs) are fundamental objects that
describe the structure of hadrons probing the distribution of momentum and spin 
among  their constituent partons. 
They also serve as an essential input for collider experiments to obtain 
the cross section of a given process. 
PDFs are extracted from global QCD analyses of worldwide
experimental data assuming certain parametrizations that  thus imply an intrinsic model dependence. 
%\cite{Perez:2012um,DeRoeck:2011na,Alekhin:2011sk,Ball:2012wy,Aidala:2012mv,Forte:2013wc,Jimenez-Delgado:2013sma,Rojo:2015acz,Butterworth:2015oua,Accardi:2016ndt,Gao:2017yyd,Lin:2017snn}.
Since PDFs are inherently nonperturbative observables,
an {\it ab-initio} calculation  using lattice 
field theory methods based solely on QCD is highly desirable. 
However, such an effort was not possible for a long time 
due to the fact that PDFs are defined on the light cone 
and thus they are not accessible by Euclidean lattice QCD simulations.

A way to circumvent this difficulty was suggested 
by Ji~\cite{Ji:2013dva} who proposed to use 
matrix elements from purely spatial correlations that are accessible in lattice QCD. In this way, 
quasi-PDFs can be connected to the true PDFs 
through a matching procedure. 
The method relies on the so-called large momentum effective theory 
(LaMET), which requires hadron states boosted to large enough momentum. Several
works implemented this proposal within lattice QCD with very promising results \cite{Lin:2014zya,Alexandrou:2014pna,Alexandrou:2015rja,Chen:2016utp,Alexandrou:2016jqi} 
demonstrating the applicability of this approach. A major step forward was 
the development of perturbative~\cite{Constantinou:2017sej} 
and nonperturbative~\cite{Green:2017xeu,Alexandrou:2017huk} renormalization 
of the lattice matrix elements.   
An overview of the current status of lattice QCD calculation of PDFs can be found 
in  Refs.~\cite{Monahan:2018euv,Cichy:2018mum}.

In Ref.~\cite{Alexandrou:2019lfo} was presented a detailed analysis of systematic effects that enter in the computation of quasi-PDFs
and need to be investigated.
One of the open issues that was identified was the 
unphysical {\em oscillations} seen in the $x$-dependence of quasi-PDFs, which also affect
the final PDFs. A possible origin of these oscillations is
 the periodicity of Fourier transformation and  the fact that a truncation is implemented due to the finite Wilson line.
 Another possible explanation is that   
not large enough momentum boosts are currently feasible. 
It is the goal of this paper 
to examine whether the discrete Fourier 
transformation is responsible   for the oscillatory behavior and to replace it by evaluating a continuous Fourier transform on the function resulting from a Gaussian process regression (GPR).
Although our approach is reminiscent  to the one described in Ref.~\cite{ambrogioni2017integral},
it differs from it in a number of aspects, as discussed in Sec.~\ref{sec:quasi-pdfs}. \par The rest of the paper is organized as follows: in Sec.~\ref{sec:dataset} we present the $N_f=2$
lattice ensemble under study in this work
analyzing the main features of phenomenological data for the unpolarized PDFs.
In Sec.~\ref{sec:DFT} we analyze in more detail the discretization
procedure related to the computation of the quasi-PDFs and the discrete Fourier transform (DFT).
In Sec.~\ref{sec:GPR} we introduce the main concepts of GPR and explain how to use it in the context of  Fourier transforms (FT).
In Sec.~\ref{subsec:mock_test}
we test the proposed method on a mock dataset generated from a function whose FT
is known in closed form.
Finally, in Sec.~\ref{sec:real_data}
we apply the Bayes-Gauss-Fourier transform
to a set of lattice data, and in Sec.~\ref{sec:conclusions} we draw our conclusions. 
\section{Quasi-PDFs and relation to PDFs}\label{sec:quasi-pdfs}

Quasi-PDFs are defined by
\begin{equation}\label{eq:quasi-pdf}
\begin{split}
    \tilde{q}(x,P_3,\mu)=2P_3\int_{-\infty}^{\infty}\frac{dz}{4\pi}&e^{ixP_3z}h(z,P_3,\mu)\\ &+\mathcal{O}\left(\frac{\Lambda^2_{QCD}}{P_3^2},\frac{M_N^2}{P_3^2}\right)
\end{split}
\end{equation}
where $M_N$ is the nucleon mass, $P_3$ is the nucleon momentum and $x$ the 
momentum fraction carried by the parton. The matrix element $h(z,P_3,\mu)$ in Eq.~\eqref{eq:quasi-pdf} reads
\begin{equation}\label{eq:matrix_el}
    h(z,P_3,\mu)=\bra{h(P_3)}\overline{\psi}(0)\Gamma W(0,z)\psi(z)\ket{h(P_3)}.
\end{equation}
In  Eq.~\eqref{eq:quasi-pdf} the state $\ket{h(P_3)}$ is the boosted hadron with four-momentum $P=(E,0,0,P_3)$ and $W(0,z)$ 
is the Wilson line taken along the boost direction $z$. The structure of the 
matrix $\Gamma$, acting in Dirac space, determines the three types of PDF, namely
$\Gamma=\gamma_3$ is for the unpolarized, $\Gamma=\gamma_5\gamma_3$ for the polarized and 
$\Gamma=\sigma_{3\nu}$ for the transversity. From finite momentum lattice QCD measurements 
of $\tilde{q}(x,P_3,\mu)$ from Eq.~\eqref{eq:quasi-pdf}, the physical PDF can be obtained through a matching procedure 
based on large momentum effective theory (LaMET) ~\cite{Xiong:2013bka,Chen:2016fxx,Wang:2017qyg,Stewart:2017tvs,Izubuchi:2018srq,Alexandrou:2015rja}. 
This allows to obtain,   
after a proper renormalization, the physical PDF, in the limit of large enough momenta.
 The matching formula reads
\begin{equation}\label{eq:matching}
    q(x,\mu)=\int_{-\infty}^{\infty}\frac{d\xi}{\abs{\xi}}C\left(\xi,\frac{\xi \mu}{x P_3}\right)\tilde{q}\left( \frac{x}{\xi},\mu,P_3 \right),
\end{equation}
where $q(x,\mu)$ is the physical PDF and $C(\xi,\eta)$ is the matching kernel that depends on the type of 
PDF. A detailed description of the matching procedure, including the analytic expression of the kernel for unpolarized PDF, is reported in Appendix~\ref{app:matching_procedure}.\\
However, on the lattice, renormalized matrix element is only available for a limited range of $z$-values, where $z$ is the length of the Wilson line entering the nonlocal operators. Since the theory is defined on a discrete set of lattice sites, the matrix element is known only for discrete values of $z$. Thus, the integral defining the quasi-PDF becomes a finite sum 
\begin{equation}\label{eq:discretization}
    2P_3 \int_{-\infty}^{\infty}\frac{dz}{4\pi}e^{ixP_3z}\rightarrow 2 P_3 \sum\limits_{z=-z_{\rm max}}^{z_{\rm max}} \frac{a}{4\pi}e^{ix P_3 z},
\end{equation}
where $a$ is the lattice spacing.
In particular, the transformation given in Eq.~\eqref{eq:discretization} means that, instead of computing the Fourier transform (FT), we compute an analytic continuation of the discrete Fourier transform (DFT) defined for continuous $x$ values. The DFT frequencies are
\begin{equation}\label{eq:omega_DFT}
    \omega_k = \frac{2\pi}{N}k, \quad k\in[-k_{\rm{max}},k_{\rm{max}}],
\end{equation}
with $N=(2k_{\rm max}/a+1)$ and $a k_{\rm{max}}=z_{\rm{max}}/a$.
However, the FT frequencies relevant for the computation of quasi-PDFs are
\begin{equation}\label{eq:omega_def}
    \omega= x P_3  = 2\pi\frac{P}{L}x , 
\end{equation}
where the momentum $P_3=2\pi P/L$, $L$ is the spatial extent of the lattice, $P\in\mathbb{N}$ and $x$ assumes continuous values in the interval $[-1,1]$. 
\\
The discretization procedure described in Eq.~\eqref{eq:discretization} introduces a  systematic bias in the quasi-PDF~\cite{Karpie:2019eiq}:
\begin{enumerate}
    \item
        knowing the matrix element only at  discrete $z$-values limits the high frequency components to $|x|<\frac{\pi}{aP_3}$.
        Therefore, higher frequency components cannot be resolved and they are wrongly measured as lower frequency components below the threshold. This phenomenon is also known as \emph{aliasing}.
    \item
    The finite spatial extent of the lattice introduces a cutoff on $z$ that is limited by $z_{\rm max}\leq \frac{L}{2}$.
    A limitation in the number of points reduces the frequency resolution of the discretized Fourier transform defined in Eq. \eqref{eq:discretization}. When frequency components of the signal do not correspond to the discrete frequencies, the discretized Fourier transform suffers from a distortive effect known as frequency \emph{leakage}.
    Cutoff effects become significant for $aP_3\sim 1$.

\end{enumerate}
These two problems could be solved if it were possible to reconstruct a continuous form of the
renormalized lattice matrix element, before evaluating the FT.
In this work we use the formalism of Gaussian process regression (GPR)~\cite{williams2006gaussian}
to perform both the interpolation and the extrapolation required in order to tackle, respectively,
the two issues mentioned above.  

The choice of GPR for the continuous reconstruction is based on the following reasons:
\begin{itemize}
    \item
        the interpolation is nonparametric, so it has the flexibility to adapt to any dataset without being restricted to a specific parametrized function;
    \item
        since GPR is based on Bayesian inference, the information about the behavior of the function toward
        infinity (extrapolation) can be incorporated into the prior distribution,
        and taken into account for the interpolation; 
    \item
        the uncertainties of the measurements are incorporated into the interpolation through Bayes theorem;
    \item
        it is possible to impose a chosen level of smoothness to the interpolating function;
    \item
        the result of the interpolation is continuous, defined over whole domain of interest and its Fourier
        transform is computable in closed form.
\end{itemize}
An application of GPR to Fourier analysis has already appeared in Ref.~\cite{ambrogioni2017integral},
where the Fourier transform of the interpolating function is referred to as Bayes-Gauss-Fourier Transform (BGFT).
Even though the main concepts remain the same, the procedure that we propose for computing the BGFT is different
from the one described in Ref.~\cite{ambrogioni2017integral}. In that paper, the prior covariance function of the
Gaussian process is determined by the inverse FT of a spectral density model that fits the DFT data.
In our case, we tune the hyperparameters of the prior covariance function using the standard maximum likelihood estimation of type II (MLE-II)
procedure described in Ref.~\cite{williams2006gaussian}, since, within this approach, we can take into account the available prior knowledge on the renormalized matrix elements.
\section{Properties of the lattice matrix element}\label{sec:dataset}
We apply the BGFT approach  to the $\mbox{M}\overline{\mbox{MS}}$ renormalized matrix element of the unpolarized PDF. 
We use the results of Ref.~\cite{Alexandrou:2019lfo} that were computed using a gauge  ensemble with two dynamical mass degenerate light twisted mass quarks ($N_f=2$) generated by the Extended Twisted Mass Collaboration (ETMC)~\cite{Abdel-Rehim:2015pwa}.  The lattice spacing is $a=0.0938(2)(3)$, the lattice size  $48^3\times 96$ and the source-sink time separation is fixed at $t_s=12a \approx 1.1\,\text{fm}$. All the relevant parameters regarding the employed gauge ensemble are reported in Table~\ref{Table:params}. This ensemble is referred to as the {\it cA2.09.48} ensemble. \par

\begin{table}[h]
  \begin{center}
    \renewcommand{\arraystretch}{1.2}
    \renewcommand{\tabcolsep}{5.5pt}
    \begin{tabular}{c|lc}
      \hline
      \hline
      \multicolumn{3}{c}{ \quad$\beta{=}2.10$,\qquad $c_{\rm SW}{=} 1.57751$,\qquad $a{=}0.0938(3)(2)$~fm\quad}\\
      \hline
      $48^3\times 96$\,  & $\,\,a\mu = 0.0009\quad m_N = 0.932(4)$~GeV   \\
      $L=4.5$~fm\,  & $\,\,m_\pi = 0.1304(4)$~GeV$\quad m_\pi L = 2.98(1)$    \\
      \hline\hline
    \end{tabular}
    \caption{\small{Simulation parameters of the {\it cA2.09.48} ensemble used to extract the unpolarized quasi-PDF. The nucleon mass $(m_N)$, the pion mass $(m_\pi)$ and the lattice spacing $(a)$ were determined in Ref.~\cite{Alexandrou:2017xwd}.}}
    \label{Table:params}
  \end{center}
\end{table}

\begin{figure}[!h]
  \includegraphics[width=\columnwidth]{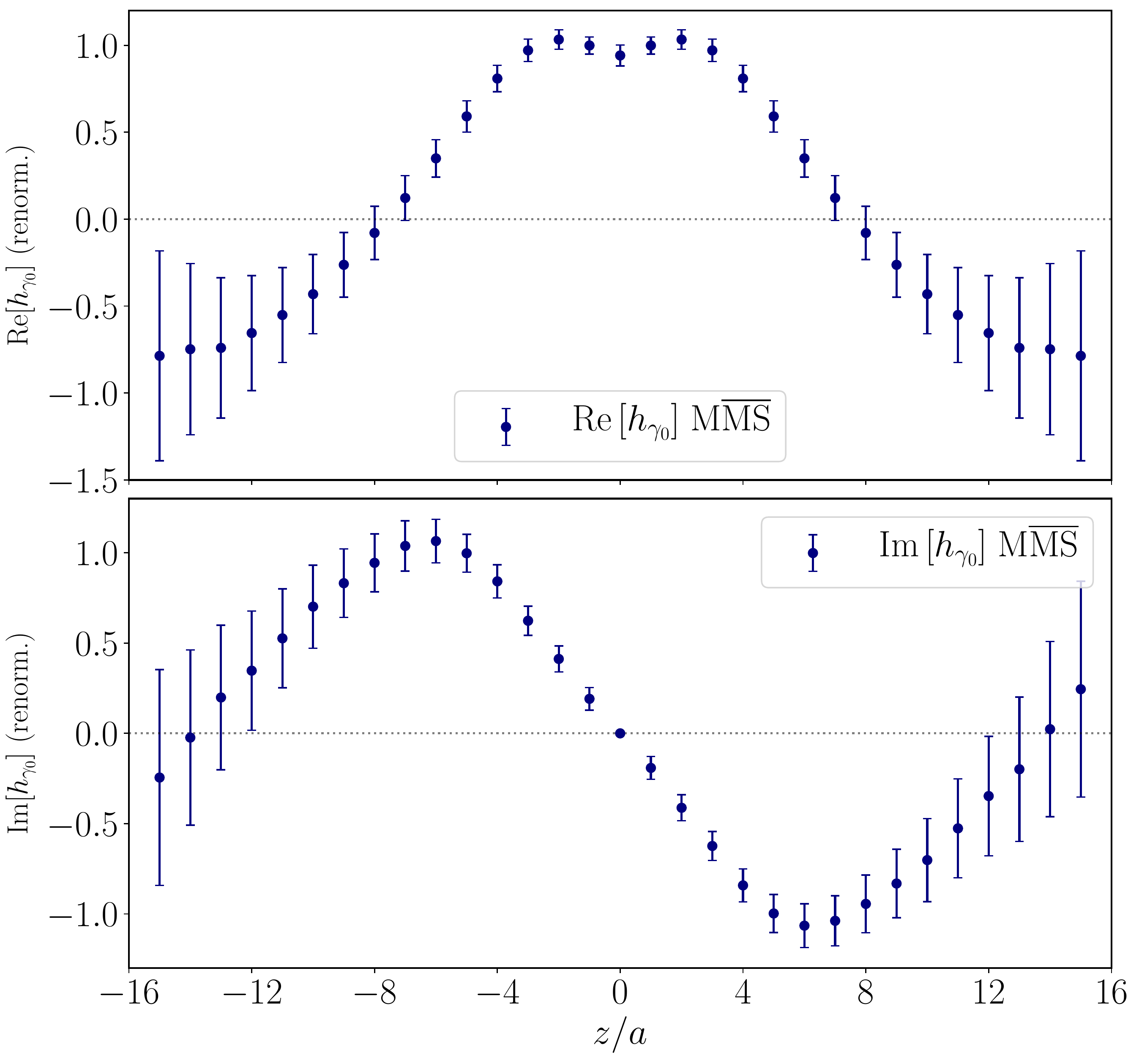}
  \caption{Real and imaginary part of the renormalized matrix element computed at $P_3 = 10 \pi/L  \simeq \ 1.38\,{\rm GeV}$.}
  \label{fig:me}
\end{figure}
We show the results on the renormalized nucleon matrix element of the unpolarized operator in Fig.~\ref{fig:me} taken from Ref.~\cite{Alexandrou:2019lfo}.  The momentum is $P_3 = 10 \pi/L  \simeq \ 1.38\,{\rm GeV}$. A total of $N_{conf}=811$ gauge configurations were analyzed, with a total number of measurements $N_{measures}= 73000$.
The errors were obtained using  the jackknife resampling method.

Before presenting the methods used to determine the quasi-PDFs, it is necessary to analyze some important aspects regarding the properties of these lattice matrix element. In a recent paper~\cite{Alexandrou:2019lfo}, based on the data from the {\it cA2.09.48} ensemble,  the authors identified possible sources of systematic uncertainties in the reconstruction of PDFs from lattice QCD simulations, including the dependence on the cutoff $z_{\rm max}$. The results showed that the PDFs extracted are contaminated by a larger noise and stronger oscillations as $z_{\rm max}$ increases above the value of $z_{\rm max}/a\gtrsim 10$. In particular, the higher the value of $z_{\rm max}$, the bigger the errors prohibiting the PDFs from reaching a zero value in the large positive $x$ region. However, all the light-cone PDFs extracted using different values of the cutoff $z_{\rm max}/a$ are compatible within errors~\cite{Alexandrou:2019lfo}. We chose to compare the results of the proposed nonparametric regression procedure with the DFT quasi-PDF obtained with $z_{\rm max}/a=10$.

Let us examine now  some important properties exhibited by the renormalized matrix element, that will be useful to better design our algorithm. First, they are Hermitian functions of $z$, namely
\begin{equation}
  h(z,P_3,\mu)=h(-z,P_3,\mu)^\dagger,
\end{equation}
which means that the real part is even and the imaginary part is odd in $z$. This property will be exploited by our BGFT method in order to make the algorithm more stable, as described in Sec. \ref{sec:hermitian}. 

Second, both the real and imaginary parts are expected to decay to zero, with a rate
that increases with the nucleon boost. This crucial aspect will be reflected in the choice of the prior mean of the GPR, which encodes all the relevant information about the renormalized matrix element, as illustrated in Sec. \ref{sec:prior_choice}.
Third, the real and imaginary parts of the matrix element can be written in polar coordinates, i.e. at fixed $z$, the complex function $h(z)=\Re h(z)+i\Im h(z)$ can be written  as
\[
h(z)=\rho(z)e^{i\phi(z)},
\]
with
\begin{equation*}
  \begin{split}
    \rho(z)&=\sqrt{\Re h(z) ^2+\Im h(z) ^2}\\
    \phi(z)&=\arg(h(z))\\
    &=\arctan\!2\left(\Im h(z),\Re h(z)\right).\\
  \end{split}
\end{equation*}
The function $\arctan\!2(y,x)$ reads
\begin{equation*}
  \begin{split}
    &\arctan\!2(y,x)=\\
    &=\begin{cases}
      2\arctan{\frac{y}{\sqrt{x^2+y^2}+x}} & \mbox{if } x>0\mbox{ or }y\neq0 \\
      \pi & \mbox{if } x<0\mbox{ and }y=0 \\
      \mbox{undefined} &\mbox{if }x=0\mbox{ and }y=0\end{cases}
  \end{split}
\end{equation*}

\begin{figure}[!h]
  \includegraphics[width=\columnwidth]{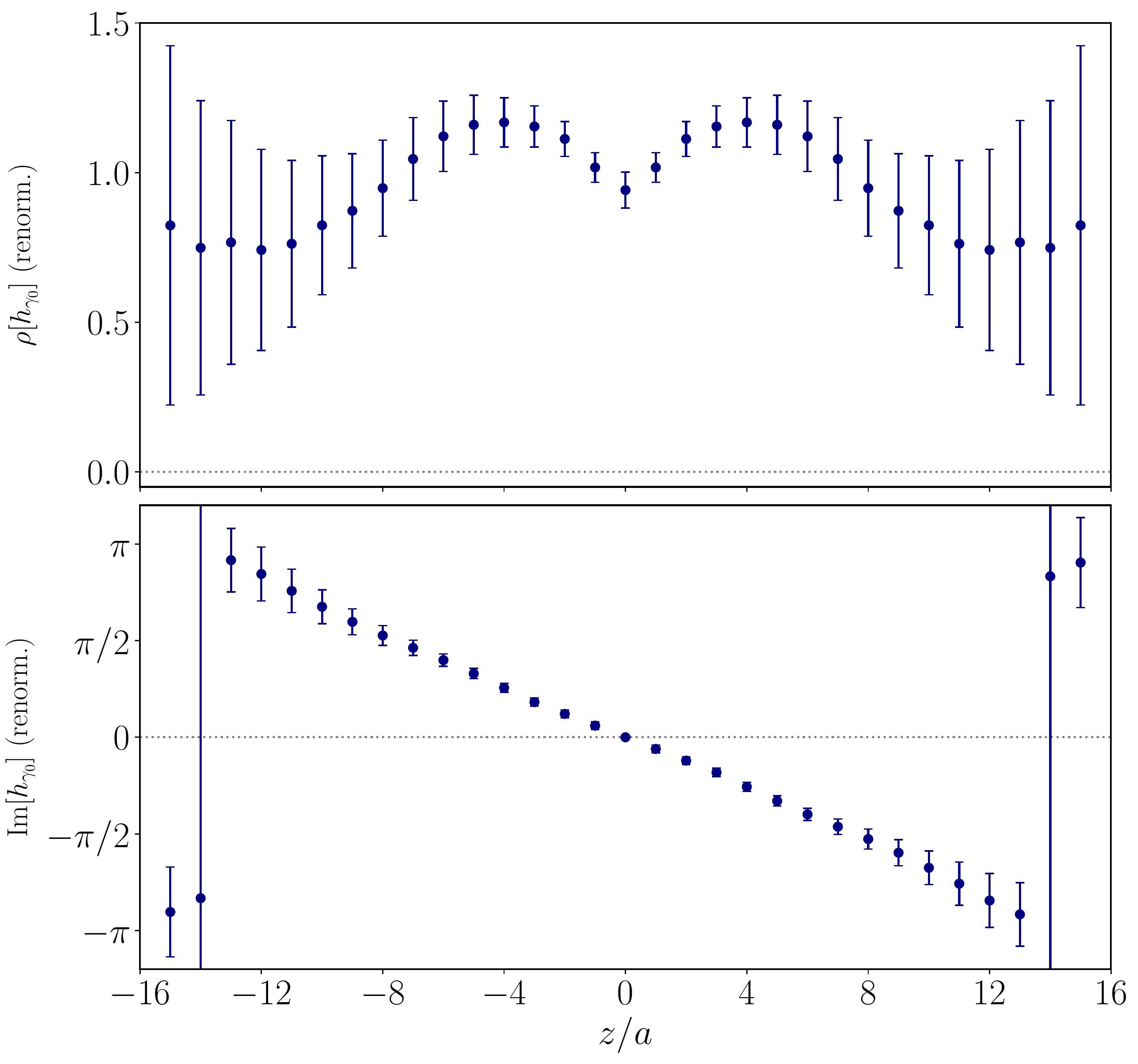}
  \caption{Modulus $\rho(z)$ and argument $\phi(z)$ of the renormalized matrix element $h(z)$.}
  \label{fig:me_rotated}
\end{figure}

In Fig.~\ref{fig:me_rotated} we show the functions $\rho(z)$ and $\phi(z)$ for the renormalized nucleon matrix element of the unpolarized operator. 
The modulus $\rho (z)$ of the complex function $h(z)$ is an even function that decays to zero with a rate that depends on the nucleon boost. However,large uncertainties in the large-$z$ region affect the asymptotic behavior  of $\rho(z)$. On the other hand, the argument $\phi(z)$ shows a linear behavior for all relevant values of $z$. Such a property means that the matrix element can be written as
\begin{equation}
  \begin{split}
    \Re h(z) &= \rho(z) \cos{\left(\theta z \right)}\\
    \Im h(z) &= \rho(z) \sin{\left(\theta z \right)},
  \end{split}
\end{equation}
or, equivalently
\begin{equation}\label{eq:h_property}
  h(z) = \rho(z) e^{i\phi(z)}=\rho(z) e^{i\theta z}.
\end{equation}
This property will be used in the BGFT to improve the results as described in Sec.~\ref{sec:analytic_ft}.
The linear dependence of $\phi(z)$ on $z$ can also be observed  directly from the phenomenological PDFs, as  discussed in the following section.

\subsection{Analysis of phenomenological data}\label{sec:phenomenological_data}
\begin{figure}[!htb]
  \includegraphics[width=\columnwidth]{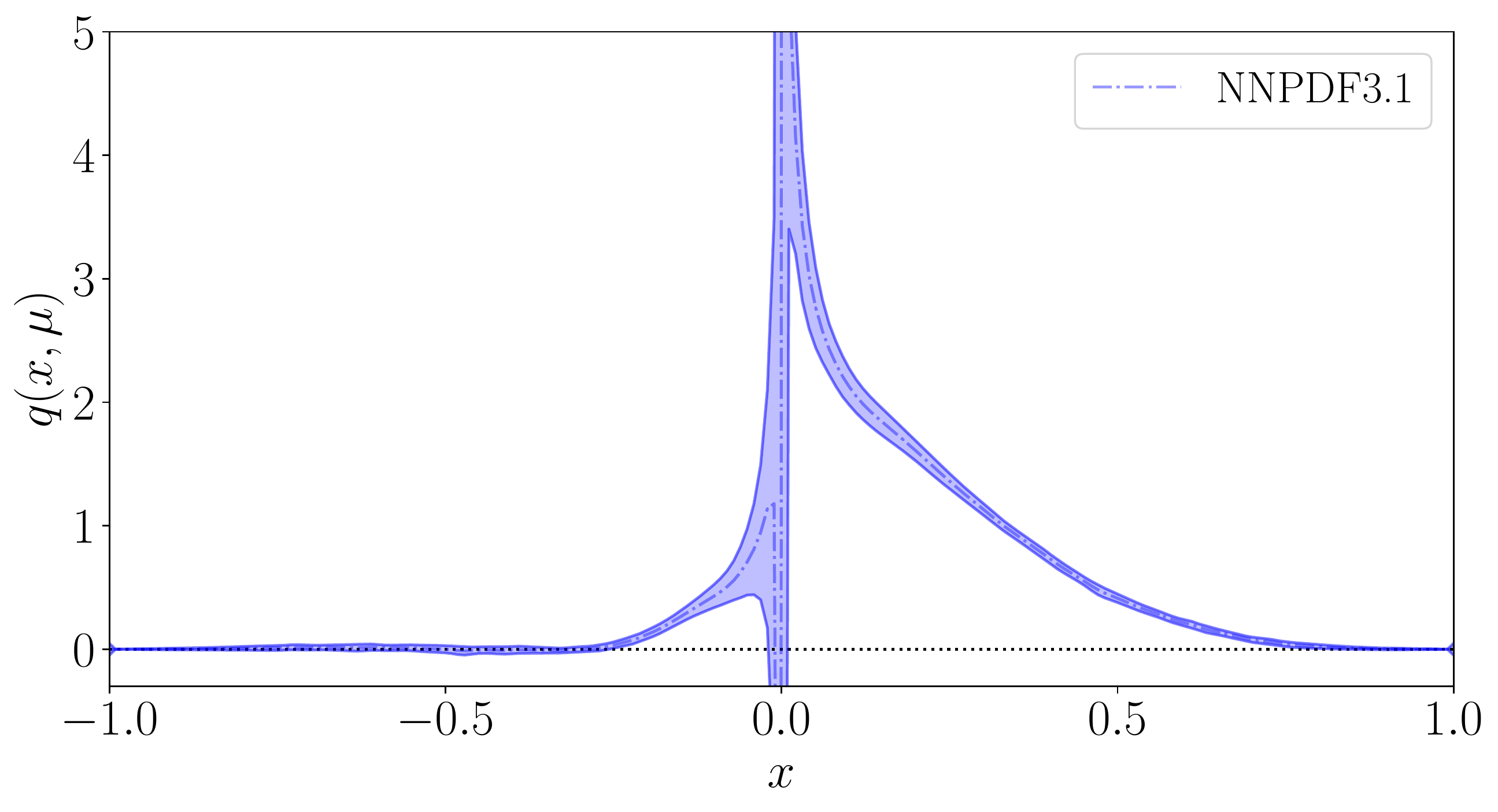}
  \caption{NNPDF3.1 phenomenological determination of the unpolarized PDF.}
  \label{fig:NNPDF31}
\end{figure}

\begin{figure}[!htb]
  \includegraphics[width=\columnwidth]{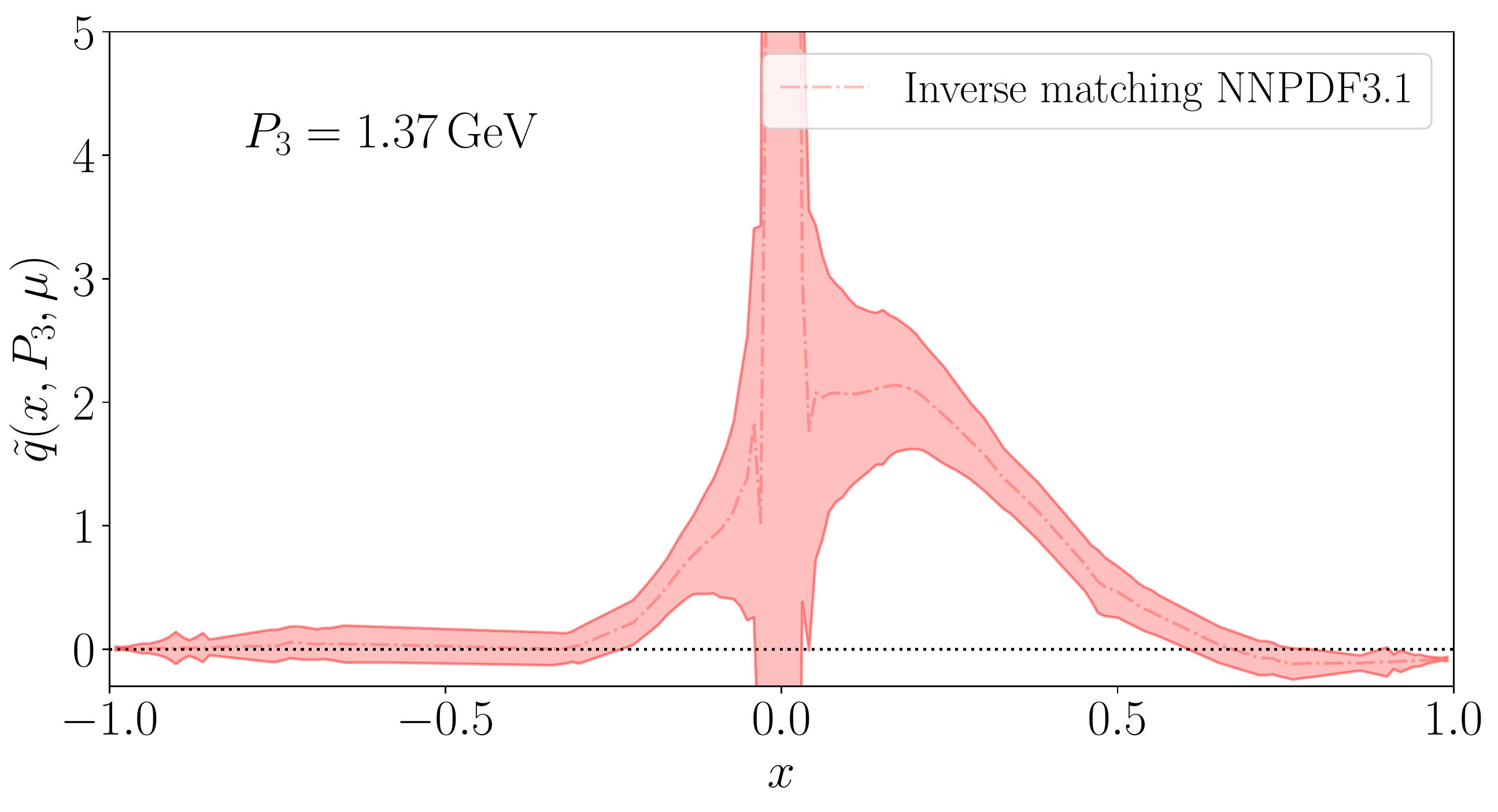}
  \caption{Quasi-PDF for $P_3=1.37\,{\rm GeV}$ obtained through the inverse matching procedure of the NNPDF3.1 data.}
  \label{fig:NNPDF31_inv}
\end{figure}

\begin{figure}[!htb]
  \includegraphics[width=\columnwidth]{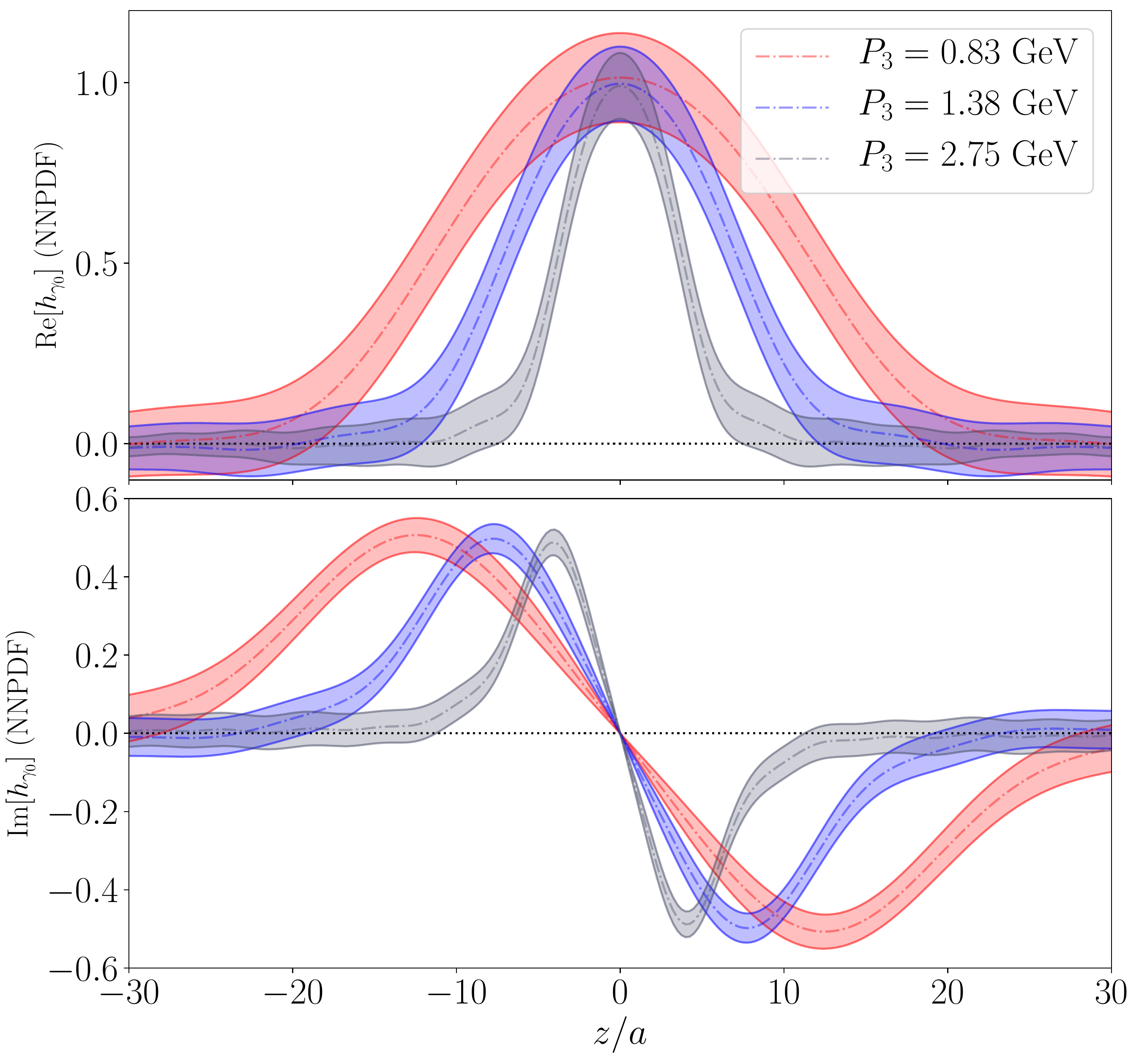}
  \caption{Matrix elements corresponding to the NNPDF3.1  phenomenological determination of the unpolarized PDF computed for three different values of the momentum $P_3$.}
  \label{fig:NNPDF31_ME}
\end{figure}

\begin{figure}[!htb]
  \includegraphics[width=\columnwidth]{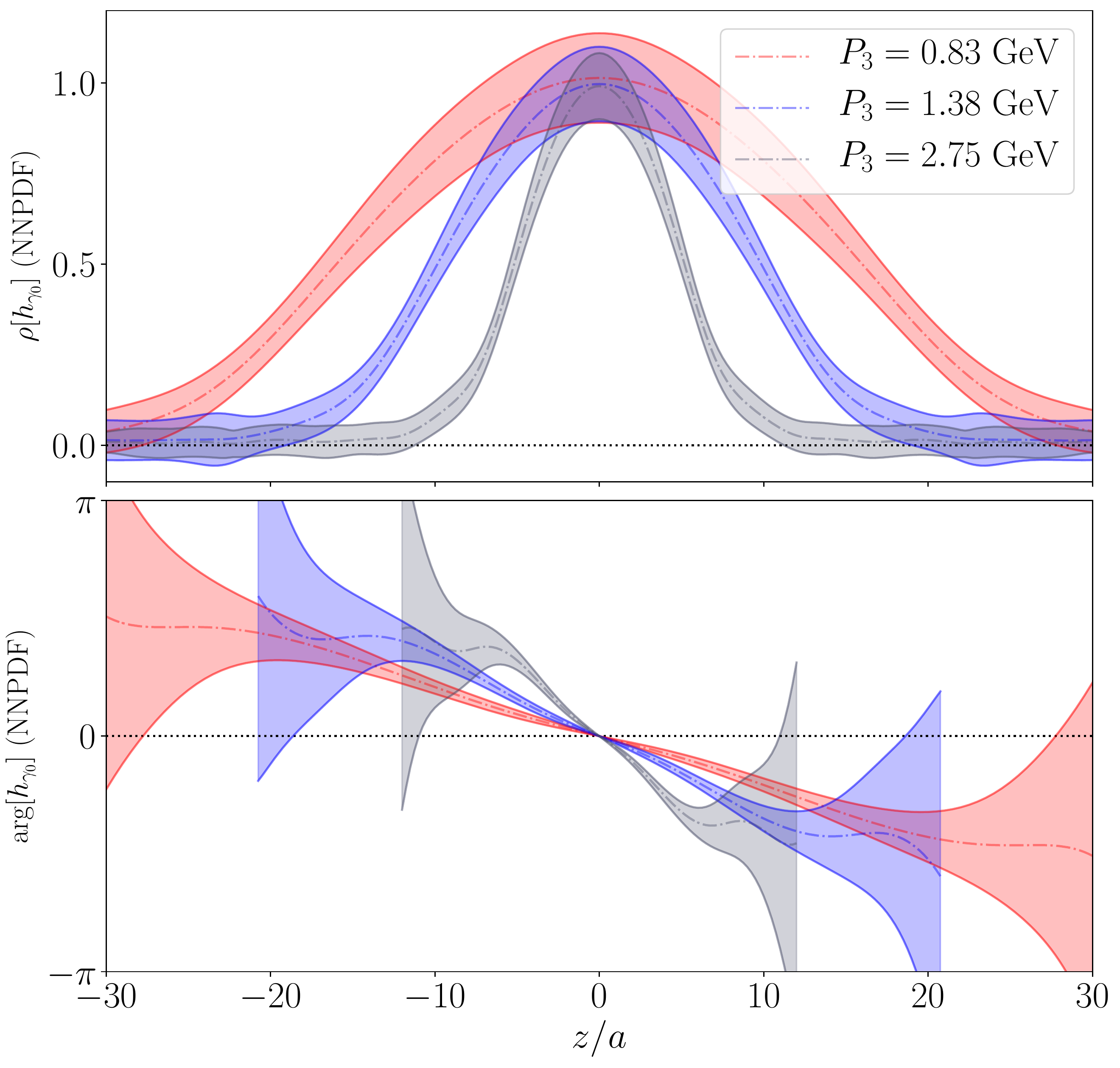}
  \caption{Modulus and argument of the matrix elements corresponding to the NNPDF3.1  phenomenological determination of the unpolarized PDF computed for three different values of the momentum $P_3$.}
  \label{fig:NNPDF31_ME_ROT}
\end{figure}

\begin{figure}[!htb]
  \includegraphics[width=\columnwidth]{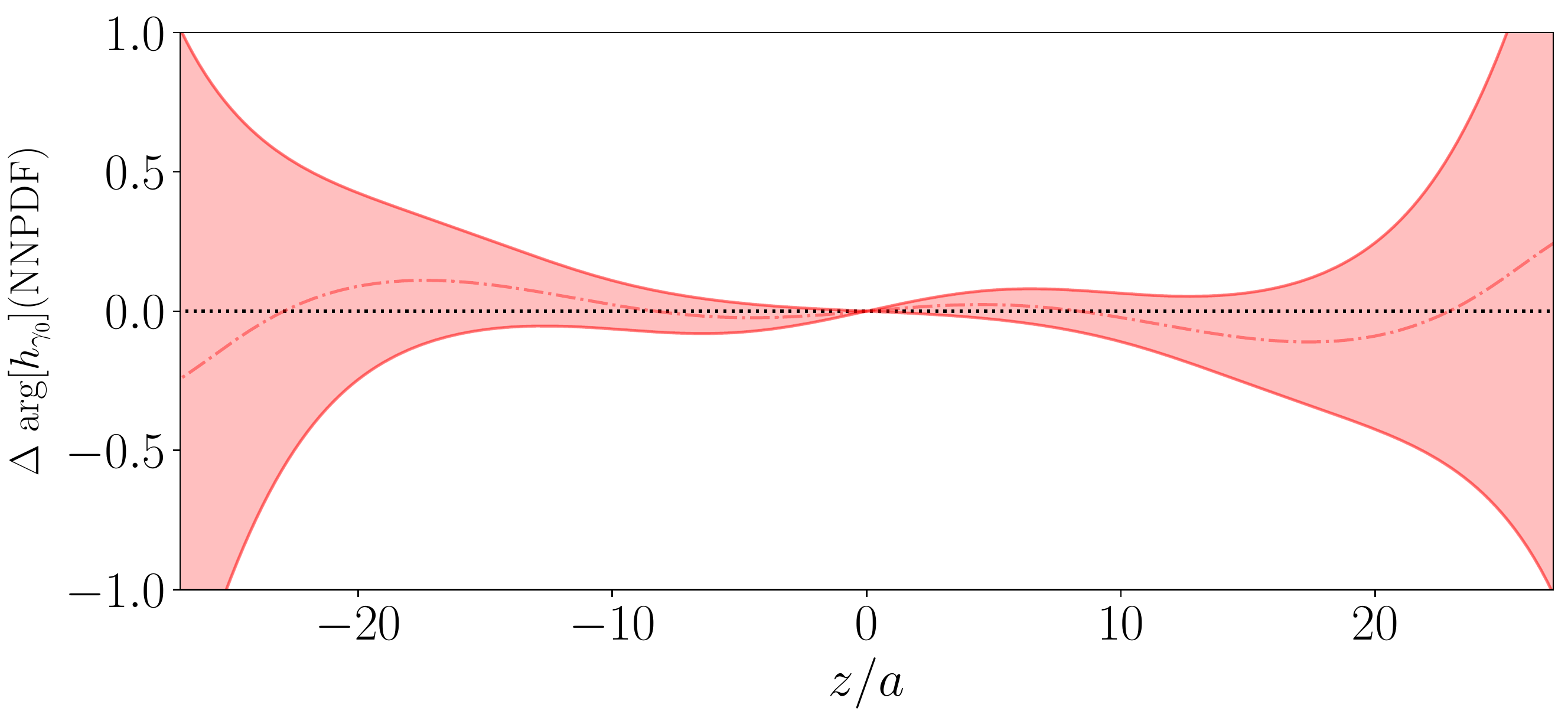}
  \caption{Deviation from the linear behavior of the argument of the matrix element corresponding to the NNPDF3.1  phenomenological determination of the unpolarized PDF computed for $P_3=0.83\,{\rm GeV}$.}
  \label{fig:NNPDF31_LIN}
\end{figure}

To illustrate the relation between the real and imaginary parts of the underlying  matrix element, we consider the NNPDF3.1 phenomenological determination of the unpolarized PDF~\cite{Ball:2017nwa}. This data set is shown in Fig.~\ref{fig:NNPDF31}  and we apply the \emph{inverse} matching procedure to derive the corresponding matrix element. The inverse matching can be interpreted as the inverse of the operation given in Eq.~\eqref{eq:matching}, which allows us to obtain the quasi-PDF from the light cone PDF by
\begin{equation} \label{eq:inverse_match_2}
  \tilde{q}\left( x,\mu,P_3 \right)=\int_{-1}^{1}\frac{dy}{\abs{y}}\,\tilde{C}\left(\frac{x}{y},\frac{\mu}{y P_3}\right)q\left( y,\mu\right),
\end{equation}
with $y=x/\xi$.The matching kernel $\tilde{C}(\xi,\eta)$ is reported in Appendix~\ref{app:matching_procedure}. The results for the quasi-PDF for $P_3=1.38\,{\rm GeV}$ are shown in Fig.~\ref{fig:NNPDF31_inv}. The larger relative uncertainties observed in the quasi-PDF compared to the phenomenological data reported in Fig. \ref{fig:NNPDF31} are due to the nonlinearity of the inverse matching procedure. In particular, the large error bar seen in the small-$x$ region in the phenomenological determination of the unpolarized PDF propagates in the full $x$-domain of the quasi-PDF. Having the quasi-PDF, the lattice matrix element can be computed through the inverse FT
\begin{equation}
h(z,\mu,P_3)=\int_{-\infty}^{\infty}dx\, e^{-izxP_3}\tilde{q}\left( x,\mu,P_3 \right).
\end{equation}
As can be seen in Fig. \ref{fig:NNPDF31_ME}, the  matrix elements extracted from the NNPDF3.1  phenomenological unpolarized PDF  show qualitatively the same behavior as that seen in the lattice QCD data. The value of the real part of the matrix element at $z=0$, which represents the net quark number, is ${\rm Re}[h(z=0)]=0.99(8)$ at $P_3=2.75\,{\rm GeV}$. The relatively large statistical error is due to the propagation of the uncertainties present in the phenomenological data in the small-$x$ region.  To get a more quantitative idea we can consider the net quark number from the light-cone PDF, namely, the numerical integral of the NNPDF3.1 determination of the unpolarized PDF, which gives  $I(PDF)\approx1.02$, while, the upper bound given by one $\sigma$  is $I(PDF+\sigma)\approx1.61$. \\
The matrix elements decay to zero at sufficient large $z/a$  with a trend that depends on the nucleon boost $P_3$. It is interesting to observe that one needs to boost to $P_3=2.75$~GeV so that the matrix element decay to zero for $z/a$-values larger than 10. Moreover, the $z$-dependent phase $\phi(z)$ behaves linearly  within the range of $z/a$-values for which  the real matrix elements is nonzero, as shown in Fig. \ref{fig:NNPDF31_ME_ROT}. In order to measure the deviation from the linear behavior of the function $\phi(z)$, we present in Fig. \ref{fig:NNPDF31_LIN} the difference between the argument of the matrix element computed at  $P_3=0.83\,{\rm GeV}$ and a linear fit performed in the interval $z/a\in[-11,11]$. The obtained curve is compatible with zero, therefore we have elements to presume that the linearity of $\phi(z)$ is indeed an intrinsic property of the matrix element that can be exploited to design an efficient regression algorithm. We will provide more details on how we exploit this property in Sec. \ref{sec:real_data}.

\section{Discrete Fourier Transform}\label{sec:DFT}

The quasi-PDF given in Eq.~\eqref{eq:quasi-pdf} is defined as the Fourier transform of the renormalized lattice matrix element of the unpolarized operator. 
As mentioned in Sec. \ref{sec:outline}, the physical PDF can be obtained from the 
quasi-PDF by applying a perturbative matching procedure. In particular, 
at one-loop order, the integral of Eq. \eqref{eq:matching} consists 
of a convolution of the quasi-PDF with a function possessing a singularity at 
$x/\xi=1$. Therefore, to compute the light-cone PDF it is crucial to obtain a 
trustworthy reconstruction of the quasi-PDF for continuous $x$-values and, 
in particular, in the region $x\in[-1,1]$. In what follows we give more details on
the issues that arise in the reconstruction of a continuous quasi-PDF 
originating from the renormalized matrix element. In particular, we  show the difficulties in accessing the small-$x$ region having available a limited amount of lattice data, corresponding to $\mathcal{O}(10)$ values of the Wilson line length $z$.\\
\begin{figure}
    \centering
    \includegraphics[width=\linewidth]{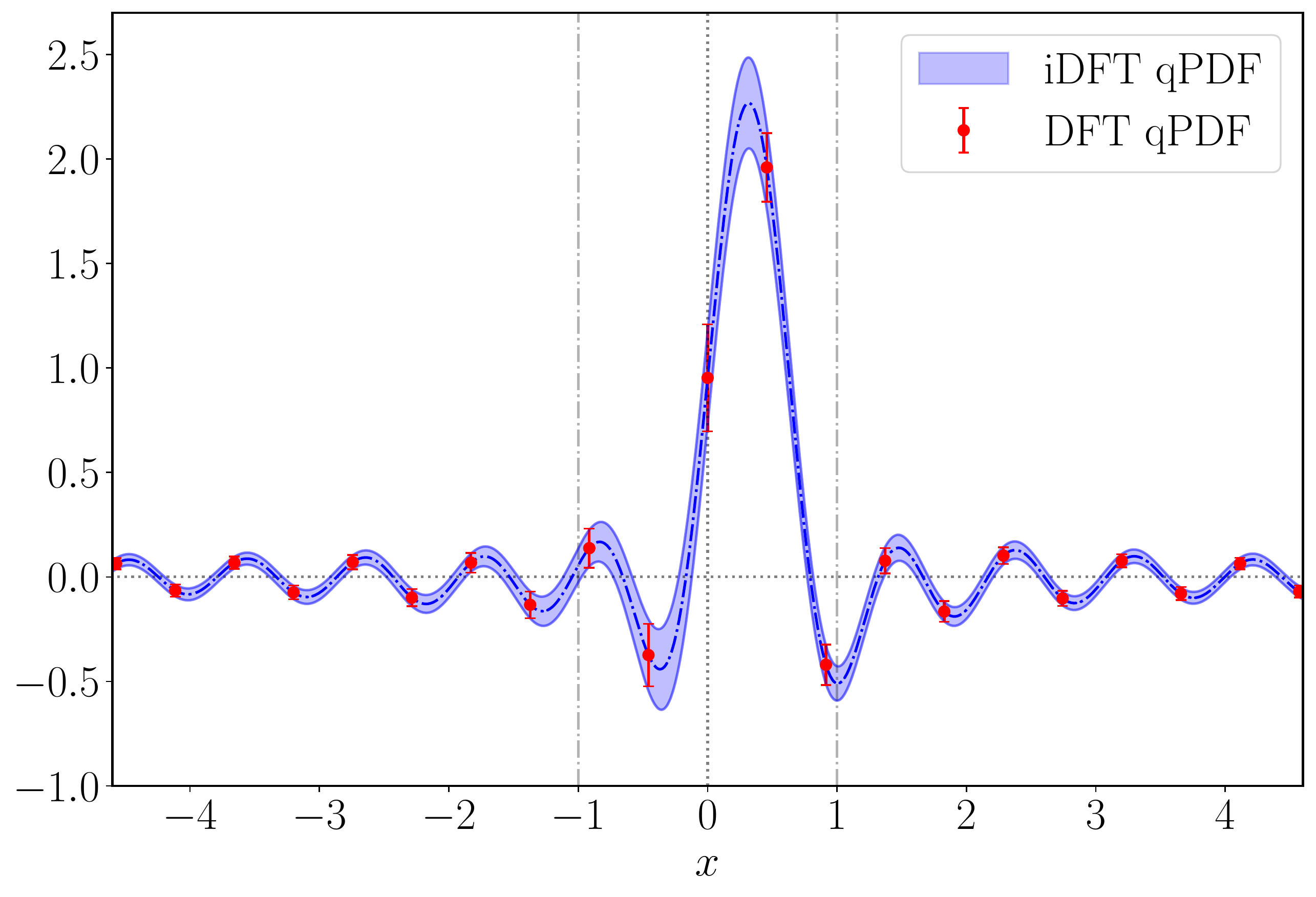}
    \caption{Comparison between DFT and iDFT corresponding to the matrix element computed on the {\it cA2.09.48}  ensemble, the parameters of which are given in Table~\ref{Table:params}.}
    \label{fig:DFT-DTFT}
\end{figure}
 The problem of reconstructing a continuous momentum-space function starting from a discrete and finite set of momentum-space points is mathematically ill-posed. Indeed, a well-defined transformation is the one that maps $N-$points discrete position-space sequence $h(z,P_3,\mu)$ defined in the interval $z\in[-z_{\rm max},z_{\rm max}]$ into a $N$-points discrete momentum-space sequence $\tilde{q}(\omega_k,\mu)$. Such a transformation is the Discrete Fourier Transform (DFT), that can be defined as follows
 \begin{equation}
     \tilde{q}(\omega_k,\mu) = 2 P_3 \sum \limits_{z=-z_{\rm max}}^{z_{\rm max}} \frac{a}{4\pi}e^{i \omega_k z} h(z,P_3,\mu),
 \end{equation}
 with 
 \begin{equation}\label{eq:DFT_freqs}
    \omega_k=\frac{2\pi}{N}k\mbox,\;\; k\in\left[-k_{\rm max},k_{\rm max}\right],
\end{equation}
where $N=(2z_{\rm max}/a+1)$ and $a k_{\rm max}=z_{\rm max}/a$.
However, the DFT  assumes periodicity of the matrix element $h(z,P_3)$ as a function of $z$ 
 \begin{equation}\label{eq:periodic_condition}
     h(z,P_3)=h(z+T,P_3),
 \end{equation}
with a period $T$ that is a function of $z_{max}$ and the DFT frequencies given in Eq. \eqref{eq:DFT_freqs} have a meaning only under this property. However,   the matrix element is not periodic and  Eq.~\eqref{eq:periodic_condition} does not hold.  In particular, this implies that we cannot attribute to the DFT frequencies of Eq.~\eqref{eq:DFT_freqs} any special meaning. However, as it will become clear later on, it is still interesting to examine  the DFT of the renormalized matrix element. In Fig.~\eqref{fig:DFT-DTFT} we show the 
DFT corresponding to the matrix element of the unpolarized operator (introduced in Sec. \ref{sec:dataset}) with a boost  
$P_3=10\pi/L$ and  $z_{\rm max}/a=10$. \\
As already pointed out, to compute the physical PDF a continuous momentum-space function $q(x,P_3)$ is required. A possible solution is to compute the sum in the right-hand side of Eq.~\eqref{eq:discretization} for continuous $x$ values. The resulting transformation can be written as follows\footnote{In signal processing, this kind of transformation is referred to as the discrete-time Fourier transform (DTFT).}
\begin{equation}\label{eq:DTFT_definition}
    \tilde{q}(\omega,P_3) = 2P_3 \sum\limits_{z=-\infty}^{\infty} \frac{a}{4\pi}e^{ix P_3 z} h(z,P_3), 
\end{equation}
with $x\in {\rm I\!R}$. Such a transformation maps an $N$-point discrete position-space sequence $h(z,P_3)$ defined on the interval $z\in[-z_{\rm max},z_{\rm max}]$ into a continuous momentum-space function $\tilde{q}(x,P_3)$. However, since the matrix element is available up to $z_{\rm max}$, the definition in Eq. \eqref{eq:DTFT_definition} cannot be applied. What we compute instead is the convolution of the matrix element with a function $\chi_I(z)$ defined by
 \begin{equation}
     \chi_I(z)=
     \begin{cases*}
        1 & if $|z|<z_{\rm max}$ \\
        0 & if $|z|>z_{\rm max}$,
     \end{cases*}
 \end{equation}
 that restricts $z$ in the interval $I=[-z_{\rm max},z_{\rm max}]$.
This procedure is equivalent to computing the sum in Eq.~\eqref{eq:DTFT_definition} within the interval $[-z_{\rm max},z_{\rm max}]$, where the matrix element is known. 
In what follows, we will show that the transform obtained through this procedure is an interpolation of the DFT and, for this reason, will be referred to as interpolated-DFT (iDFT). The results of applying the iDFT to the matrix element as compared to the DFT are shown  in  Fig.~\eqref{fig:DFT-DTFT}.
The iDFT is the transformation that is being employed to compute the quasi-PDF starting from the renormalized matrix element and, for this reason, it is interesting to analyze in depth the features characterizing this quantity. \\
The matrix element can be expressed as the inverse DFT
\begin{equation}\label{eq:inverseFT}
    h(z,P_3)=\frac{4\pi}{2aP_3}\frac{1}{2N+1}\sum\limits_{k=-k_{\rm max}}^{k_{\rm max}}e^{-i\omega_k z}\tilde{q}(\omega_k,P_3),
\end{equation}
where $\tilde{q}(\omega_k,P_3)$ is the $k$th DFT coefficient, and $k_{\rm max}=z_{\rm max}/a$. Substituting Eq.~\eqref{eq:inverseFT} into Eq.~\eqref{eq:DTFT_definition}, and computing the sum over $z$ we get
\begin{equation}\label{eq:conversion_formula}
    \tilde{q}(\omega,P_3)= \sum\limits_{k=-k_{\rm max}}^{k_{\rm max}} \tilde{q}(\omega_k,P_3) D_n(\Delta \omega_k),
\end{equation}
where $\Delta \omega_k = \omega-\omega_k$ and 
\begin{equation}
    D_n(x)=\frac{1}{n}\frac{\sin{\left(x\, n/2\right)}}{\sin{\left(x /2\right)}}
\end{equation}
is the so-called Dirichlet kernel. In particular, in the limit $x \to 0 $, the Dirichlet kernel converges to $1$, thus the iDFT is equivalent to the DFT for $\omega=\omega_k,\;k\in[-k_{\rm max},k_{\rm max}]$. \\
In summary, given an $N-$point discrete position space sequence $h(z,P_3,\mu)$, it is not possible to define an appropriate discrete transformation returning an $N$-point discrete momentum-space sequence $\tilde{q}(\omega_k,\mu)$.

The most straightforward solution is to use  analytical continuation for real values of the variable $x$ of the DFT. In particular, the iDFT is the convolution of a step function $\chi_I(z)$ defined in the interval $[-z_{\rm max},z_{\rm max}]$ with the matrix element. This is equivalent to setting the value of the matrix element to zero for $|z|>z_{\rm max}$ and, as it will be shown in Sec. \ref{subsec:mock_test}, this procedure introduces nonphysical oscillations in the quasi-PDF.

\section{Bayes-Gauss-Fourier transform} \label{sec:GPR}
\subsection{Gaussian process regression}
A Gaussian process (GP)~\cite{williams2006gaussian} $f(z)$ is a collection of random variables, labeled by the continuous real index $z$, that follows a joint multivariate Gaussian distribution with mean function $\mu(z)$ and covariance function $k(z,z')$. This means that, given any number $n$ of domain points $z_1, ..., z_n$, the random variables $f_1, ..., f_n$, where $f_i\equiv f(z_i)$, are distributed according to:
\begin{equation*}
  p(f_1, ..., f_n) = \det(2\pi K)^{-1/2}e^{-\frac{1}{2}\sum_{ij}(f_i-\mu_i)K_{ij}^{-1}(f_j-\mu_j)}
\end{equation*}
where $\mu_i\equiv \mu(z_i)$ and $K^{-1}_{ij}$ is the inverse matrix of $K_{ij}\equiv k(z_i,z_j)$.

Given the values of the matrix element $h(z_1), ..., h(z_n)$, the aim of GPR is to find the most suitable $\mu(z)$ and $k(z,z')$ such that
$\mu(z)$ is an estimate of $h(z)$, and the covariance $k(z,z')$ describes the deviations of $\mu(z)$ from $h(z)$.
If the matrix element $h(z_1), ..., h(z_n)$ has zero error, then an appropriate GPR would return a mean function such that $\mu(z_i)=h(z_i)$ and $k(z_i,z_i)=0\ \forall i$. For this reason, the resulting GP is also called a surrogate model for $h(z)$.

\subsection{Bayesian inference}
The regression is usually performed with Bayesian inference,
for which it is needed to specify a prior GP that is then adapted to the measured data
computing the conditional probability and using Bayes theorem.
The resulting posterior GP is the surrogate model.

In order to specify the prior GP $f_P(z)$, it is sufficient to define a prior mean function $\mu_P(z)$ and a prior covariance function $k_P(z,z')$.
The prior mean function represents our belief about the behavior of $h(z)$. The prior covariance function quantifies the amount of expected deviation from the mean function, and the correlation between those deviations at different values of $z$.

Denoting by $z_1, ..., z_n$ the $z$ values in which the matrix element measures $h_1, ..., h_n$ are available with uncertainties $\Delta h_1, ..., \Delta h_n$ and assuming the errors to be Gaussian, it is possible to analytically compute the mean function $\mu(z)$ of the posterior Gaussian process $f(z)$ in terms of the prior GP~\cite{williams2006gaussian}:
\begin{equation}\label{eq:regression}
  %\begin{cases}
  \mu(z) = \mu_P(z)+
  \sum_{ij}k_P(z,z_i)\widetilde{K}_{ij}^{-1}(h_j-\mu_P(z_j))% \\[.7em]
\end{equation}
where $\widetilde{K}_{ij} = k_P(z_i,z_j)+\Delta h_i^2\delta_{ij}$.

\subsection{Choice of the prior}\label{sec:prior_choice}
It is possible to verify from Eq.~\eqref{eq:regression} that, if the errors $\Delta h_i$ are zero, then $\mu(z)$ passes through all measured points.
This property, together with smoothness conditions that can be imposed on GP, as it will be  detailed below, ensures that the dependence of the mean posterior on the mean prior becomes weaker as we get closer to measured points. Therefore, the choice of GP prior does not affect the domain areas with high density of measured points. On the other hand, for $z$ values that are far from the measured values, the choice of the prior plays a decisive role.

With the choice of the prior GP, it is possible to impose different levels of smoothness to a GP \cite{adler1981geometry,abrahamsen1997review}.
In particular, choosing $\mu_P(z), k_P(z,z') \in C^\infty$ guarantees that both the GP prior and the GP posterior are infinitely mean-square differentiable, which we expect to be the most appropriate description for our problem at hand.

In order to not compromise the performance of the method and to reduce  overfitting,
the prior covariance function is commonly chosen to be stationary and symmetric so that $k_P(z,z')=k_P(|z-z'|,0)$.
Another common choice is to use a covariance function monotonically decreasing with distance:
$k_P(a,0)>k_P(b,0)\ \forall a,b$ such that $0<a<b$.
This last property states that the correlation between predictions at different $z_1$ and $z_2$
decreases with their distance $|z_1-z_2|$, which means that the value of $\mu(z)$
will be mostly determined by the value of the neighboring measured points.

We opted for the squared exponential covariance function, which is the standard choice for a $C^\infty$ function that satisfy all the properties listed above. We thus consider
\begin{equation}\label{eq:covariance}
  k_P(z,z') = \sigma^2e^{-\frac{(z-z')^2}{2\ell^2}},
\end{equation}
where the real values $\sigma$ and $\ell$, also called hyperparameters, are fixed using the maximum likelihood estimation of type II described in Ref.~\cite{williams2006gaussian}.

Since the behavior of the mean function tends to be independent of the measured values at long distances from them, the asymptotic behavior depends only on the choice of the prior mean function.
As mentioned in Sec.~\ref{sec:dataset}, we know that renormalized matrix element should tend to zero in the limit of $|z/a|\to\infty$.
It is then possible to guarantee this limit for the posterior mean function by 
choosing a prior mean function that tends to zero at infinity.

\subsection{Strategy for complex Hermitian data}\label{sec:hermitian}
The GPR described until now is defined on real-valued data. There are a number of possible 
ways to extend the procedure to complex data.  The most direct way is to split the complex signal $h(z)$ into the real and imaginary parts, and perform two independent GPRs. Denoting as $\mu^{\Re}(z)$ and $\mu^{\Im}(z)$ the two resulting GPRs,  the complex fit result of this \emph{Cartesian} decomposition is given by:
\begin{equation}\label{eq:cartesian_fit}
h^\text{fit}_\text{car}(z)=\mu^{\Re}(z)+ i\mu^{\Im}(z).
\end{equation}
However, this method does not exploit the property of Eq.~\eqref{eq:h_property} described in Sec. \ref{sec:dataset}.
Therefore, we chose an alternative approach, which fits separately the absolute value $\rho(z)$ and the argument $\phi(z)$ of $h(z)$. For the absolute value we use a GPR, while, for the complex argument of $h(z)$, we perform a minimum 
$\chi^2$ linear regression. The advantage of this method is to restrict the space of possible fitting functions in order to improve the stability and the performance of the algorithm. In particular, replacing one of the two GPRs with a linear regression, not only reduces the computing time, but also the final variance of the fitted function. This is because  a nonparametric method has more freedom in selecting the fitting function, and, thus, the final result is more likely to exhibit wider variations from the mean.

All datasets considered in this work show linearity of the complex argument, and all the presented results are obtained with the $polar$ decomposition that exploits this property. However, the linearity of the argument might not  necessarily be a general property for other values of $P_3$ and $\mu$. In those cases, one can still apply the more general, albeit less precise,  \emph{Cartesian} decomposition.

There is still a property that can be exploited by the fitting algorithm. Since our target functions are Hermitian, we can restrict the fit procedure to the positive semiaxis and then use the Hermitian symmetry to obtain
the results on the negative semiaxis.
With this strategy there is a reduction by half\footnote{ Excluding the central point at $z=0$.} of the number of points used for the nonparametric regression,
which improves the stability and the performance of the algorithm.
Since the complex argument is an odd function of $z$,
the linear regression is performed with the intercept fixed to zero.

Denoting by $\mu(z)$ the result of the GPR for the absolute value and $\theta$ the coefficient resulting from the linear regression, the surrogate model obtained for the renormalized matrix element in the positive semi axis reads:
\begin{equation*}
  h_\text{pos}^\text{fit}(z) = \mu(z)e^{i\theta z} \qquad \text{for}\ z\geq0.
\end{equation*}
The corresponding Hermitian function defined over the full real domain is then:
\begin{equation}\label{eq:hermitian_fit}
  h^\text{fit}(z) = \mu(|z|)e^{i\theta z}.
\end{equation}
A side effect of this procedure is the loss of continuity of the derivative at $z/a=0$. However this is not an issue for our procedure because no subsequent passages require this property to hold.

\subsection{Analytic Fourier transform} \label{sec:analytic_ft}
A useful feature of GPR is the possibility to perform analytically the FT of the posterior mean, obtaining  an improved stability and performance compared to what is achievable with a numerical integration.

The FT definition that we adopt is the following:
\begin{subequations}\begin{gather}
    \mathcal F[f(z)](x) \equiv \frac{1}{2\pi}\int_{-\infty}^\infty dz\,e^{ixz}f(z) \label{eq:ft}\\
    \mathcal F^{-1}[F(x)](z) \equiv \int_{-\infty}^\infty dx\,e^{-ixz}F(x) \label{eq:anti-ft}
\end{gather}\end{subequations}
If $\mathcal T$ is the integral transform defined by Eq. \eqref{eq:quasi-pdf}, it is possible to write $\mathcal T$ and $\mathcal T^{-1}$ in terms of the FT of Eq.~\eqref{eq:ft} as follows
\begin{subequations}\begin{gather}
    \mathcal T[h(z)](x) = P_3\mathcal F[h(z)](P_3x) \label{eq:transform}\\
    \mathcal T^{-1}[H(x)](z) = \mathcal F^{-1}[H(x)](P_3z) \label{eq:anti-transform}
\end{gather}\end{subequations}
Thus, after  performing the fit, it is possible to estimate the quasi-PDF by computing the FT of the fit function of Eq.~\eqref{eq:hermitian_fit} using the convention of Eq.~\eqref{eq:ft}, and then by evaluating it using Eq.~\eqref{eq:transform}.

When computing the FT, the phase of Eq. \ref{eq:hermitian_fit} simply corresponds to a shift in the FT:
\[
\mathcal F[h^\text{fit}(z)](x) = \mathcal F[\mu(|z|)](x+\theta)
\]
In order to compute the FT of $\mu(|z|)$, it is useful to observe that Eq.~\eqref{eq:regression} is just a linear combination of the covariance function reported in Eq.~\eqref{eq:covariance}:
\[
\mu(|z|) = \mu_P(|z|) + \sum_i w_ik_P(|z|,z_i),
\]
where $w_i \equiv \sum_j \widetilde{K}_{ij}^{-1}(h_j-\mu_P(z_j))$ and $\mu_P(z)$ is a generic prior mean assigned to the absolute value of the renormalized matrix element.

The FT of $k_P(|z|,z_i)$ is available in closed form given by
\begin{multline*}
  \mathcal F[k_P(|z|,z_i)](x) \\
  =  \frac{e^{-ixz_i-\ell^2x^2/2}}{2\sqrt{2\pi}/\ell}
  \bigg(1+e^{2ixz_i} + 
  \text{erf}\left(\frac{z_i/\ell-i\ell x}{\sqrt{2}}\right)\\
  +e^{2ixz_i}\text{erf}\left(\frac{z_i/\ell-i\ell x}{\sqrt{2}}\right)\bigg).
\end{multline*}
Thus, the quasi-PDF transform of the fit is 
\begin{multline}\label{eq:fit_transform}
  \mathcal T[h^\text{fit}(z)](x) = P_3\mathcal F[\rho_P(|z|)](P_3(x+\theta)) \\+ P_3\sum_i w_i\mathcal F[k_P(|z|,z_i)](P_3(x+\theta))
\end{multline}
If the chosen prior mean $\rho_P(|z|)$ has a known analytical FT, then the quasi-PDF transform of the fit is available in closed form.

Similarly, an analytic expression of the FT can be found also if one employs the \emph{Cartesian} decomposition, or in the case of a non-Hermitian target function.

\begin{figure}
  \includegraphics[width=\columnwidth]{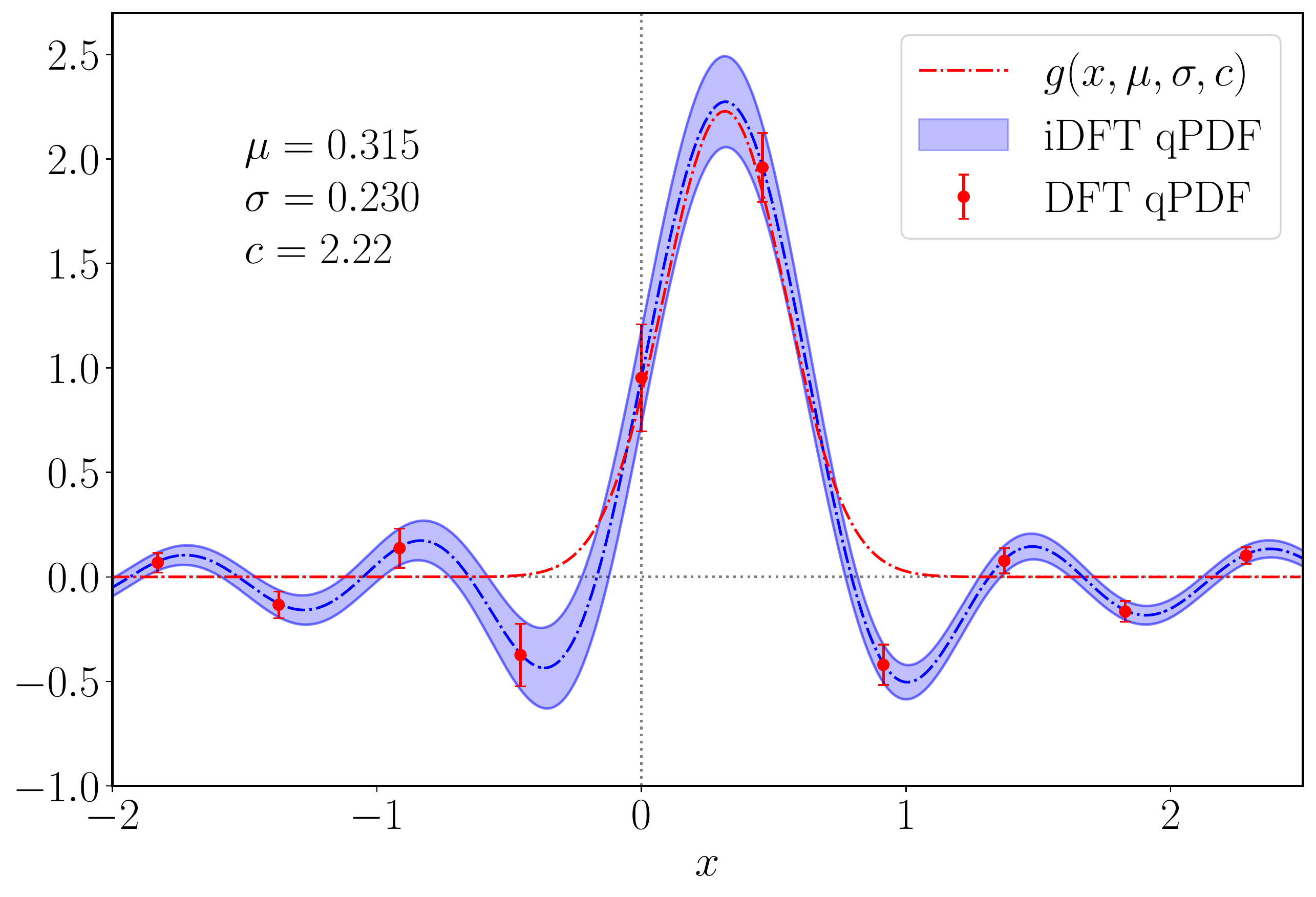}
  \caption{The function $g(x;\mu,\sigma,c)$ (line dashed line) fitted on the iDFT data extracted using the grid of points (red circles).}
  \label{fig:fit_DTFT}
\end{figure}

\section{Testing BGFT on a mock dataset}\label{subsec:mock_test}
\begin{figure}
  \includegraphics[width=\columnwidth]{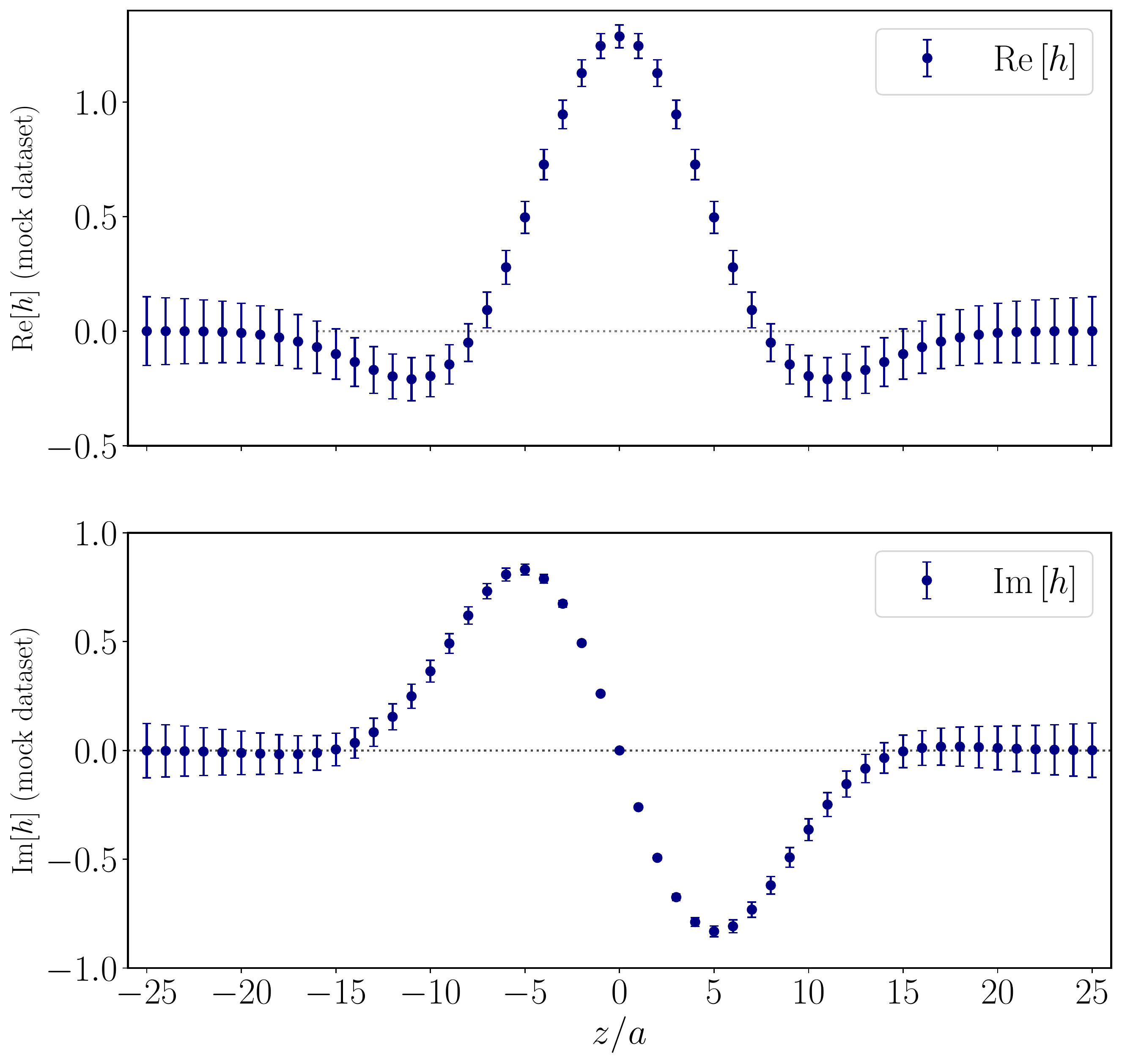}
  \caption{Mock matrix element generated according to Eq. \eqref{eq:FT_shift_gauss}.}
  \label{fig:mock_dataset}
\end{figure}

\begin{figure}
  \includegraphics[width=\columnwidth]{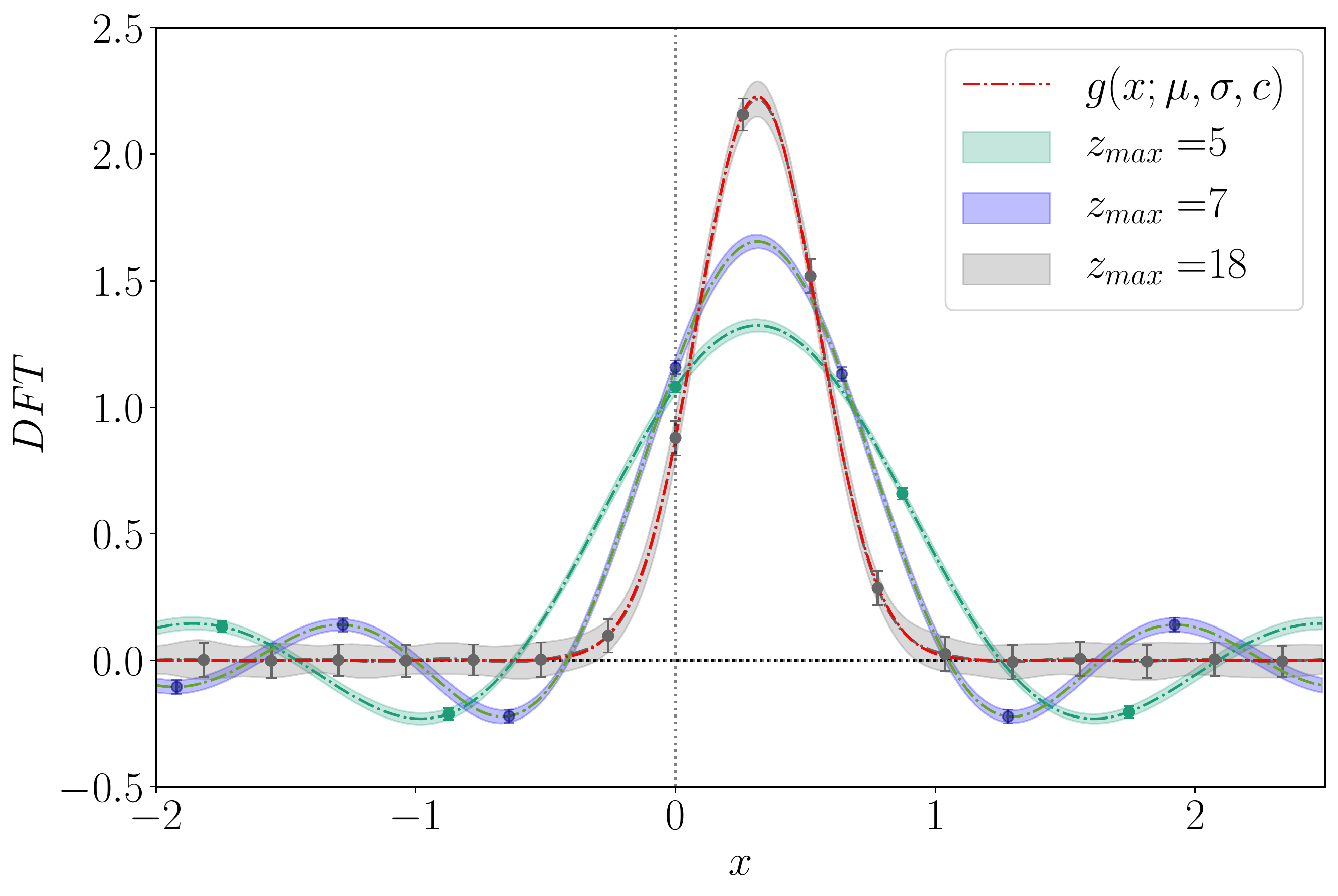}
  \caption{Dependence of the DFT on the cutoff compared to the shifted Gaussian $g(x,\mu,\sigma,c)$ from which we generated the mock dataset.}
  \label{fig:cutoff_effects}
\end{figure}

To gain an insight on the artifacts that may be introduced by the discrete Fourier transform and on the effectiveness of the proposed method, we produced a mock dataset that mimics the behavior of the matrix element shown in Fig. \ref{fig:me}. Given the rescaled Gaussian 

\begin{equation}\label{eq:shift_gauss}
  g(x;\mu,\sigma,c)=c\,\exp{-(x-\mu)^2/{2\sigma^2}}
\end{equation}
its inverse quasi-PDF transform of Eq. \eqref{eq:anti-transform} reads
\begin{equation}\label{eq:FT_shift_gauss}
  \mathcal T^{-1}[g(x)](z)=c\,\sqrt{2\pi}\sigma e^{-\sigma^2P_3^2z^2/2+i\mu P_3z}   
\end{equation}

To be consistent with the results reported in Sec. \ref{sec:real_data}, we choose 
$P=5$ and $L=48$. The complex function of Eq.~\eqref{eq:FT_shift_gauss} is
 then sampled in the interval $z\in[-25,25],\,z\in\mathbb{N}$, with 
$c=2.22$, $\mu=0.315$, $\sigma=0.230$. The employed coefficients $\mu,\sigma$ and 
$c$ correspond to the best fit performed with the function of
Eq.~\eqref{eq:shift_gauss} on the data obtained by evaluating the iDFT using the discrete set of grid points. The resulting fit is shown in Fig.~\ref{fig:fit_DTFT}.  
In order to mimic the behavior of lattice data, we generated $N=100$ numbers at each fixed integer $z$ from a Gaussian distribution centered in $\mathcal T^{-1}[g(x)](z)$ with variance increasing linearly with $z^2$, obtaining a sample of $N$ mock matrix element. The average and the Jackknife standard deviation of this sample are shown in Fig.~\ref{fig:mock_dataset}.

\begin{figure}
  \includegraphics[width=\columnwidth]{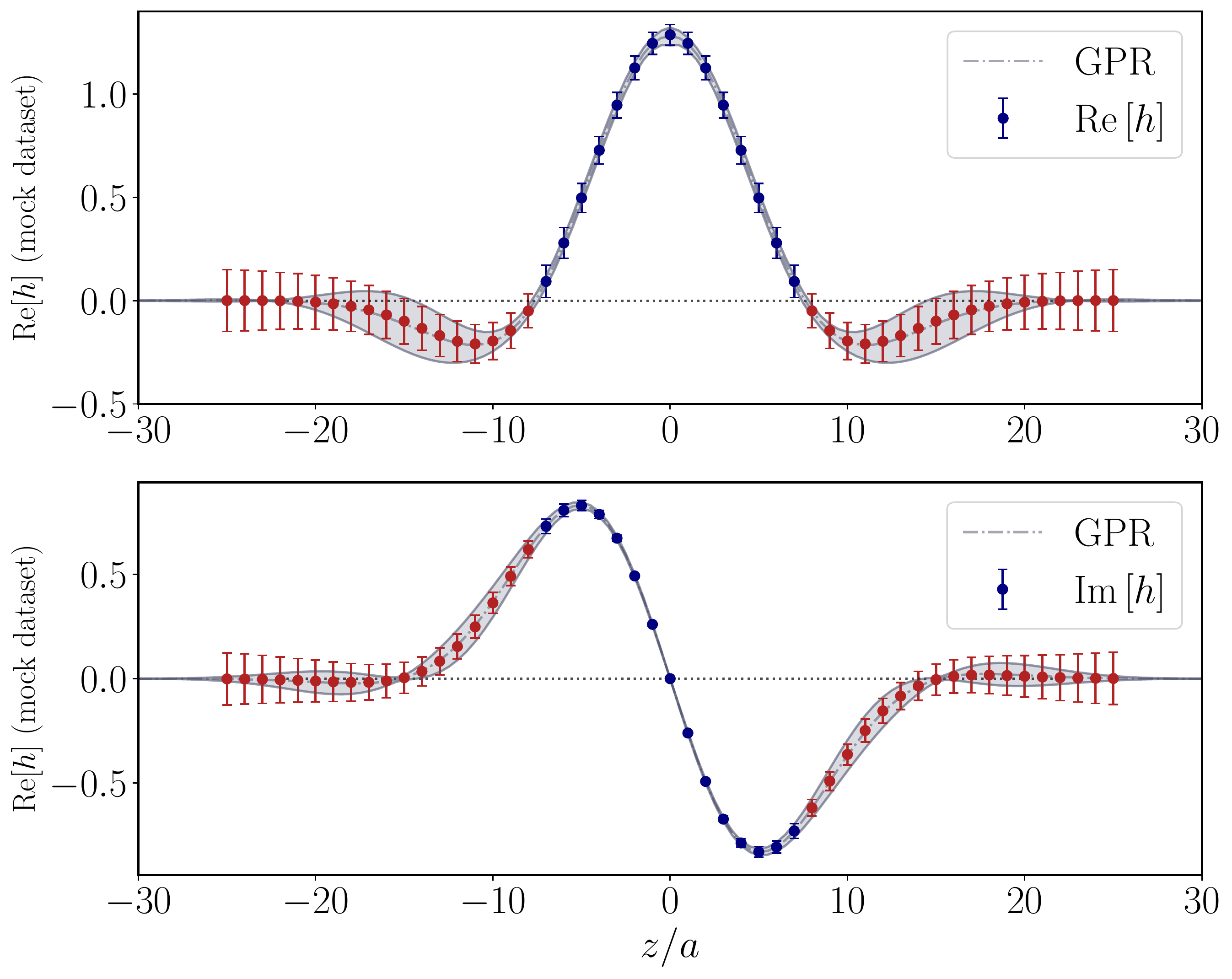}
  \caption{nonparametric regression performed on the mock data set. Only data up to $z_{\rm max}=7$ (blue circles)  are included in the fit.}
  \label{fig:GPR_example}
\end{figure}

\begin{figure}
  \includegraphics[width=\columnwidth]{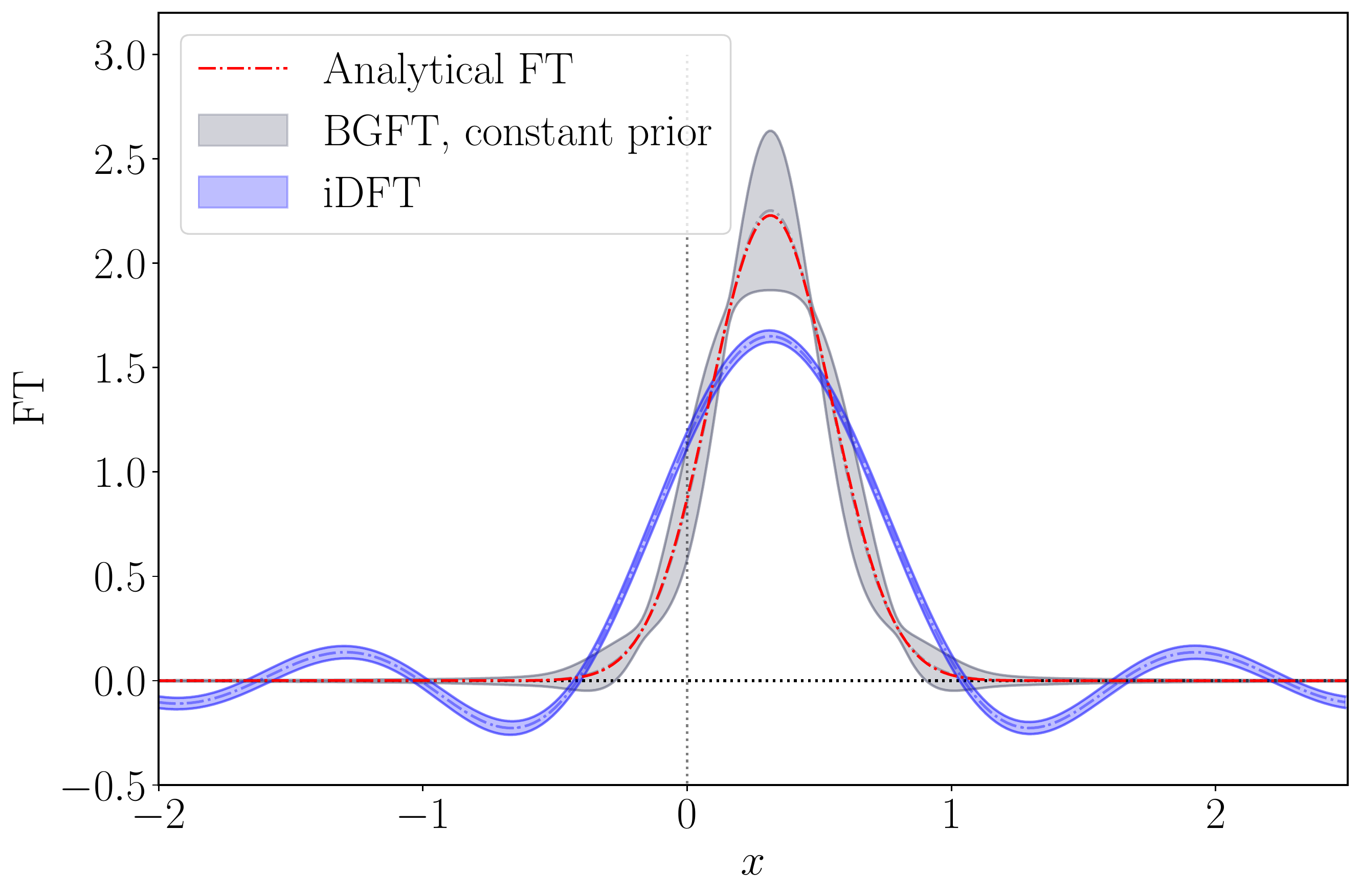}
  \caption{BGF transform performed on the mock data set and comparison with the analytical FT (red dashed line)  and the DFT (blue band).}
  \label{fig:comparisonFT}
\end{figure}

The dependence of the discrete Fourier transform on the cutoff $z_{\rm max}$ is investigated. In Fig.~\ref{fig:cutoff_effects} we show the DFT computed with four different values of $z_{\rm max}$, together with the shifted Gaussian $g(x,\mu,\sigma,c)$ from which we generate the mock dataset. In particular, setting $z_{\rm max}=5$, a huge bias is introduced in the DFT and big oscillations afflict the final result. Moreover, the bias becomes negligible only with $z_{\rm max}=18$, where the iDFT coincides with the analytical FT within the error. The observed behavior of the iDFT is due to the fact that, if the $z$ cutoff is  too small, then the frequency resolution of the DFT is not fine enough to capture the behavior of the analytical FT in the small $x$ region. Considering the case of $z_{\rm max}=7$, we apply the regression described in Sec.~\ref{sec:GPR} to the mock data set, using a constant zero function as the prior mean function.
The results are illustrated in Fig.~\ref{fig:GPR_example}. The conclusion is that, given the mock matrix element up to $z_{\rm max}=7$  and their asymptotic behavior specified by the prior as a zero constant function, the nonparametric regression is able to reproduce the data outside the fit range.  As a consequence, the results, after applying the Bayes-Gauss-Fourier transform shown in Fig. \ref{fig:comparisonFT}, are compatible with the analytical transform of the mock matrix-element within error.


\section{Application of BGFT to the renormalized lattice matrix element} \label{sec:real_data}
We apply the method detailed in Sec.~\ref{sec:GPR} on the case of interest, namely the renormalized nucleon matrix element of the unpolarized operator, for which the data are given in Sec.~\ref{sec:dataset}.
As pointed out in Sec.~\ref{sec:prior_choice}, in the region with high density of data-points, the posterior mean is strongly dependent on the renormalized matrix element rather than on the prior mean.

However, as discussed in Sec.~\ref{sec:dataset},  the matrix elements at $z_{\rm max}/a \gtrsim 7$  carry large statistical errors, and, for this reason, they are excluded from the regression. As a consequence of this, the outcome of the posterior mean become increasingly  more dependent on the prior mean function starting from $|z/a|>7$. The asymptotic behavior is almost entirely determined by the choice of the prior mean function. Therefore, any theoretical consideration on the asymptotic behavior should be incorporated into the prior mean function.

We choose to test our method with two different prior mean functions, namely the uniformly zero function and the function of Eq.~\eqref{eq:FT_shift_gauss} obtained through applying the inverse transform to the Gaussian fit of the DFT. Both of the chosen priors satisfy the requirement of being asymptotically zero.
Using different prior distributions is a method to cross-check that the final conclusions are independent of the prior choice.

\begin{figure}
    \includegraphics[width=\columnwidth]{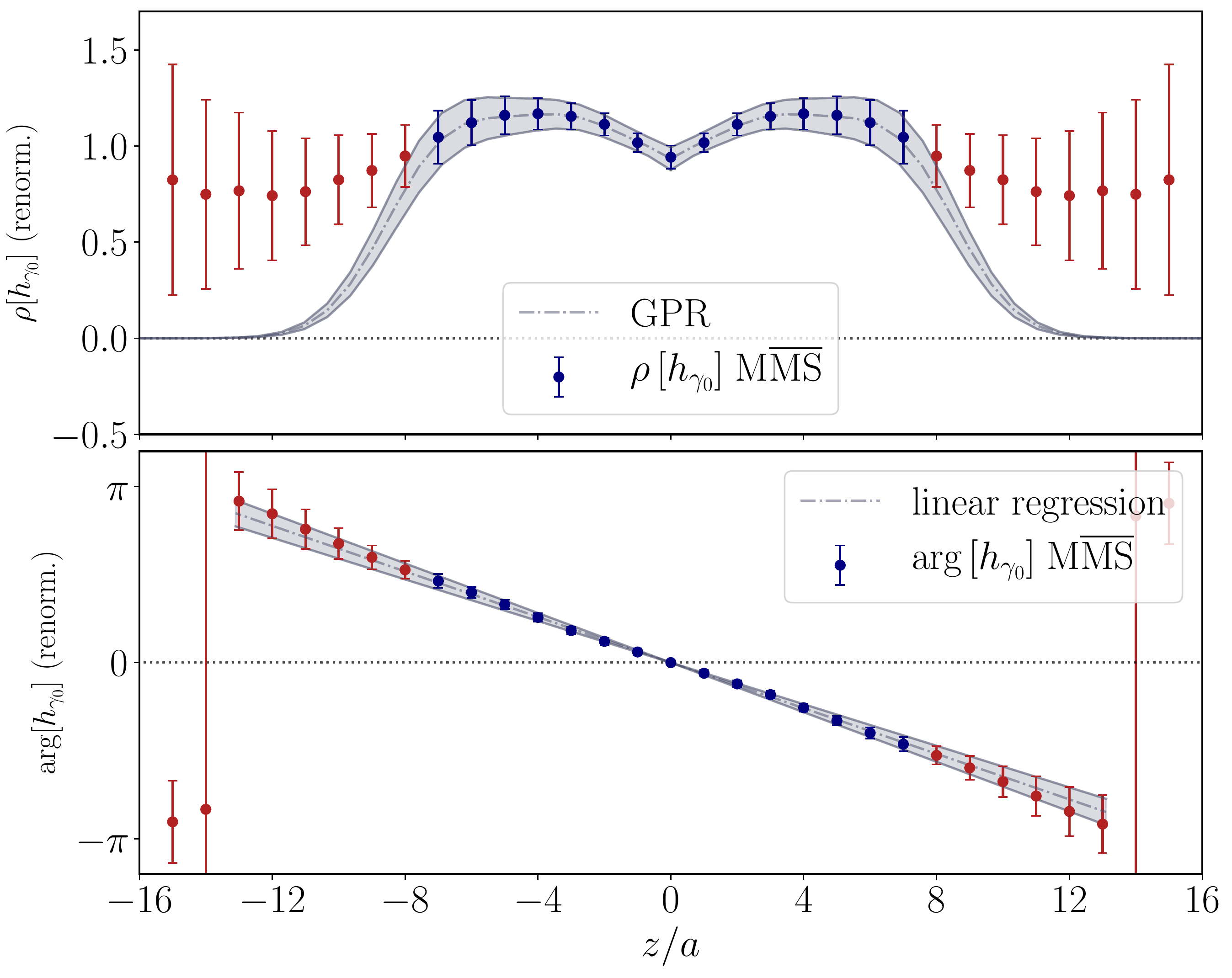}
    \caption{Lattice data for the matrix element rotated in the polar complex plane as described in Sec.~\ref{sec:real_data}. Only the points shown with the  blue circles are considered in the fit. }
    \label{fig:GPR_zeroP_polar_rot}
\end{figure}

\begin{figure}
    \includegraphics[width=\columnwidth]{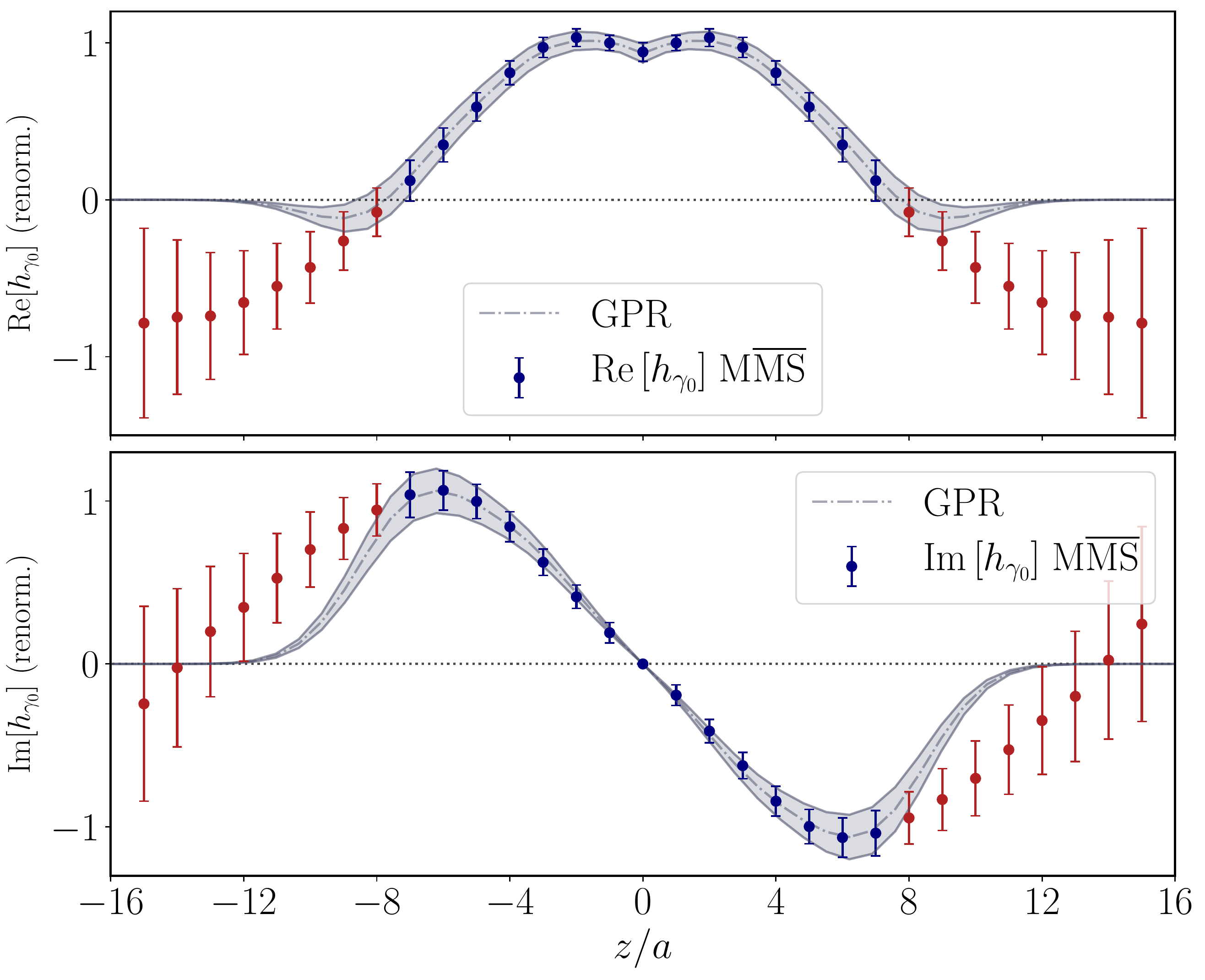}
    \caption{Lattice data for the matrix element together with the non parametric regression.  Only the points shown with the  blue circles are considered in the fit.}
    \label{fig:GPR_zeroP}
\end{figure}

Let us here summarize the key steps of the procedure:
\begin{enumerate}
    \item  At a fixed $z$-value we rewrite the complex number $h(z)=\Re h(z)+i\Im h(z)$ in the polar complex plane as
    \[
    h(z)=\rho(z)e^{i\phi},
    \]
    with
    \begin{equation*}
    \begin{split}
        \rho(z)&=\sqrt{\Re h(z) ^2+\Im h(z) ^2}\\
        \phi(z)&=\arg(h(z))\\
        &=\arctan\!2\left(\Im h(z),\Re h(z)\right).\\
    \end{split}
    \end{equation*}
    The function $\arctan\!2(y,x)$ is defined in Sec.~\ref{sec:dataset}
    \item As pointed out in Sec.~\ref{sec:dataset}, the function $\rho(z)$ is asymptotically zero, while $\phi(z)$ can be taken as a linear function of $z$.
      After choosing a prior mean function, we perform a non parametric regression of the function $\rho(z)$, while a linear fit is sufficient to reproduce $\phi(z)$, as shown in Fig.~\ref{fig:GPR_zeroP_polar_rot}; 
    \item In order to check the result of the fit to the renormalized matrix element, 
we can go back to the Cartesian coordinates, as shown in Fig.~\ref{fig:GPR_zeroP};
    \item  Employ the formula in Eq. \eqref{eq:fit_transform} to compute the quasi-PDF.
\end{enumerate}

 \begin{figure}
    \includegraphics[width=\columnwidth]{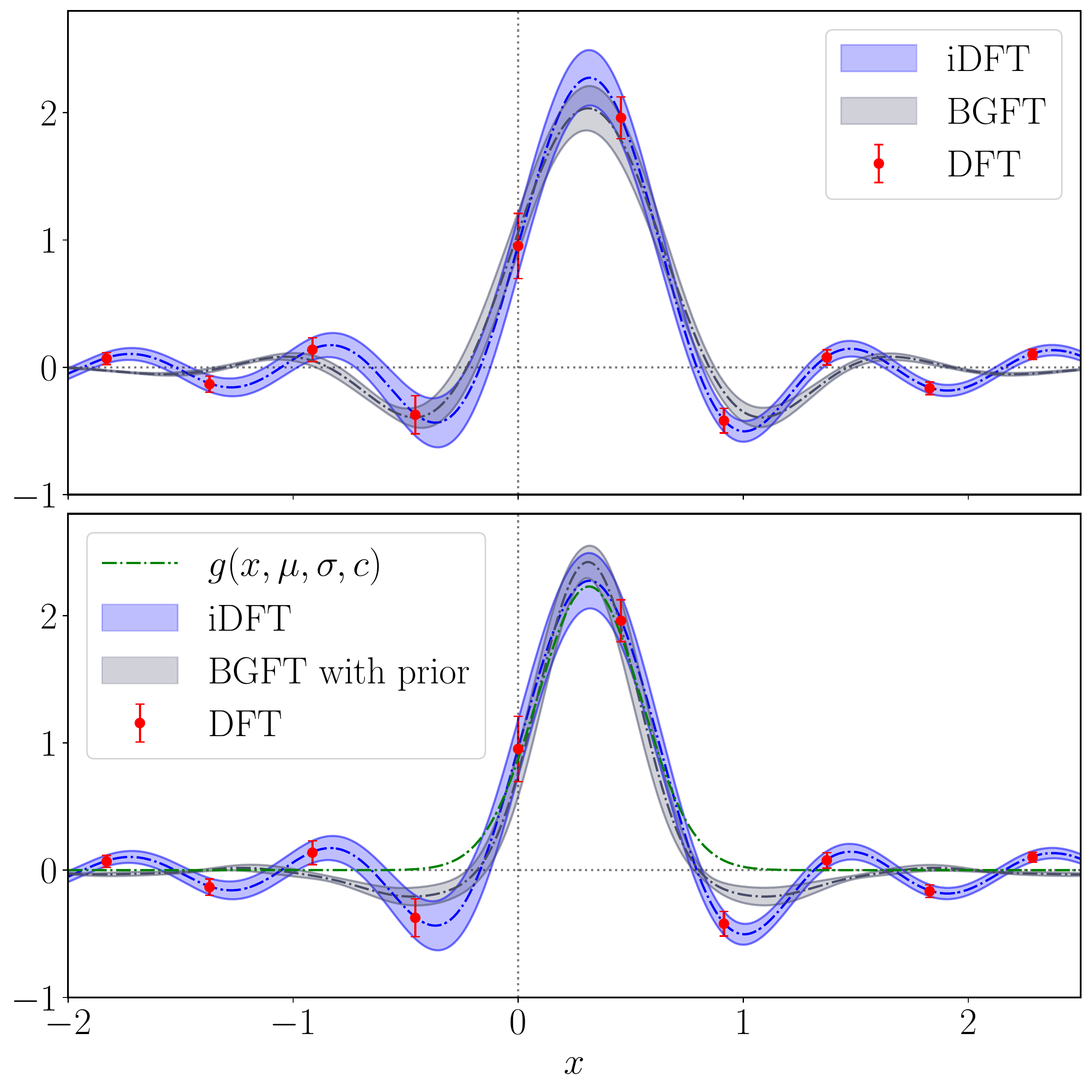}
    \caption{Unpolarized quasi-PDF computed with discrete Fourier transform (red circles), iDFT (blue band) and Gauss-Bayes-Fourier transform (grey band) with zero mean prior (upper panel) and with mean prior $g(x,\mu,\sigma,c)$ (lower panel).}
    \label{fig:qPDF_comparison}
\end{figure}
\begin{figure}
    \centering
    \includegraphics[width=\linewidth]{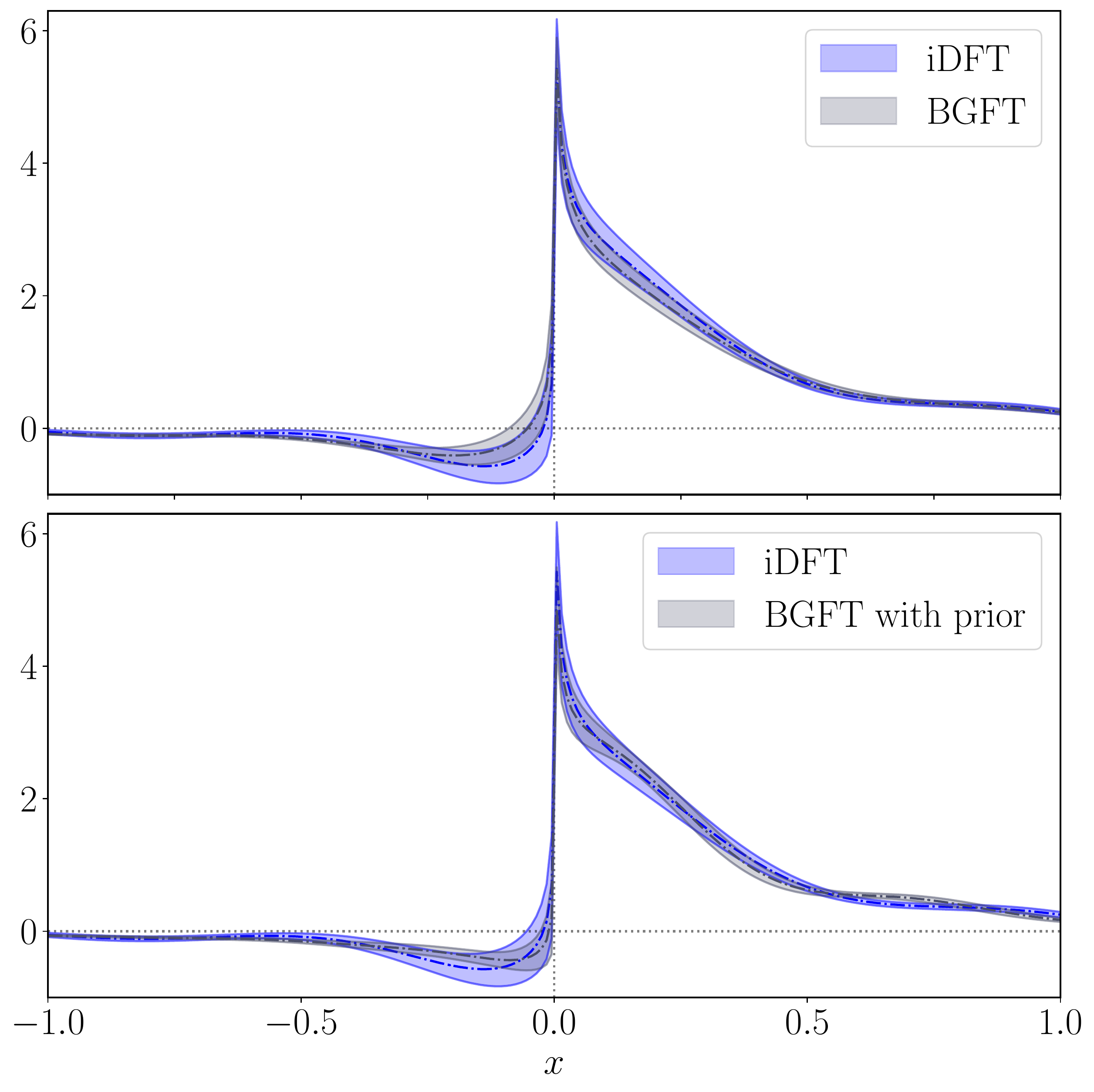}
    \caption{Comparison of light-cone PDFs obtained via iDFT reconstruction and BGFT with zero mean prior (upper panel) and with mean prior $g(x,\mu,\sigma,c)$ (lower panel).}
    \label{fig:matched_comparison}
\end{figure}

In the upper panel of Fig.~\ref{fig:qPDF_comparison} we compare the iDFT quasi-PDF to the BGFT quasi-PDF. While in the physical region $x\in[-1,1]$ the two results are compatible, for larger $\abs{x}$ nonphysical oscillations due to the periodicity of the discrete FT are strongly suppressed. However, the physical meaning of quasi-PDF can be made explicit only after having performed the matching procedure. In Fig.~\ref{fig:matched_comparison} we display the light-cone PDF reconstructions obtained via iDFT and BGFT. As can be seen,  although the  the nonphysical oscillations in the quasi-PDF are suppressed, in the physical PDF the effect is small. The  nonphysical negative PDF in the antiquark region $x\sim -0.1$ remains, as well as a mild oscillatory behavior in the large $\abs{x}$ region. This means that this behavior does not appear to be caused by the cutoff in $z$ and the discrete FT.

\begin{figure}
    \centering
    \includegraphics[width=\linewidth]{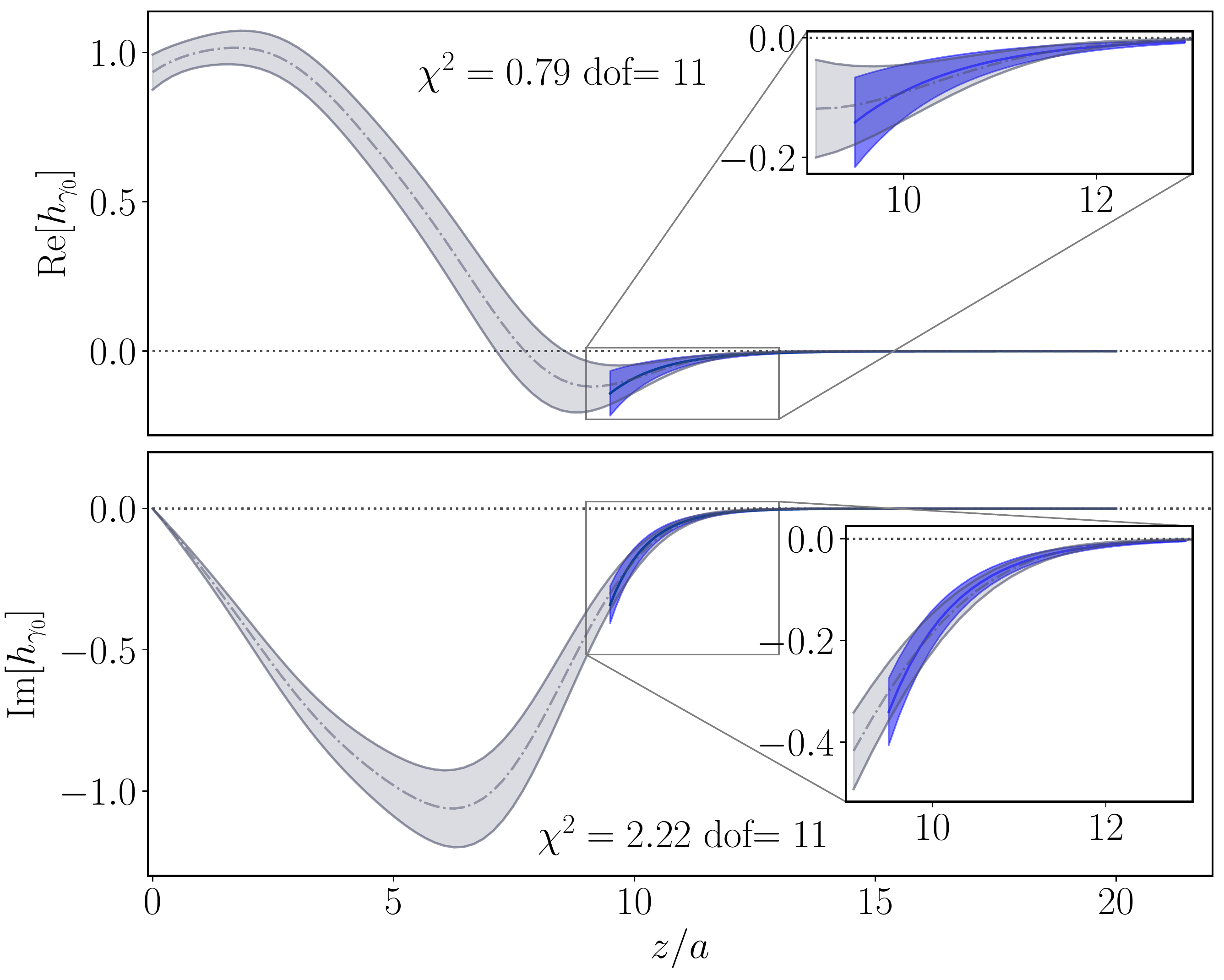}
    \caption{Fit (blue bands) with an exponential model of the tails of the nonparametric regression curves (grey bands) corresponding to the real (upper panel) and imaginary (lower panel) part of the matrix elements. }
    \label{fig:fit_tails}
\end{figure}

Finally, it is interesting to investigate how the nonparametric regression curves for the real and imaginary parts of the matrix element go to zero. In particular, both the tail of the real and imaginary parts of the nonparametric regression curves can be modeled by the function
\begin{equation}\label{eq:exp_model}
    s(z) = -a^2\exp{-zb}.
\end{equation}
The parameters $a$ and $b$ have been computed by minimizing  $\chi^2$. We obtained $a=25(16),\;b=9.4(9)\;{\rm fm}^{-1}$ for the real part and $a=270(80),\;b=13.8(7)\;{\rm fm}^{-1}$ for the imaginary part. In Fig.~\ref{fig:fit_tails} , we show the result of the fit procedure performed with the exponential model of Eq.\eqref{eq:exp_model}.

\par
\begin{figure}
    \includegraphics[width=\columnwidth]{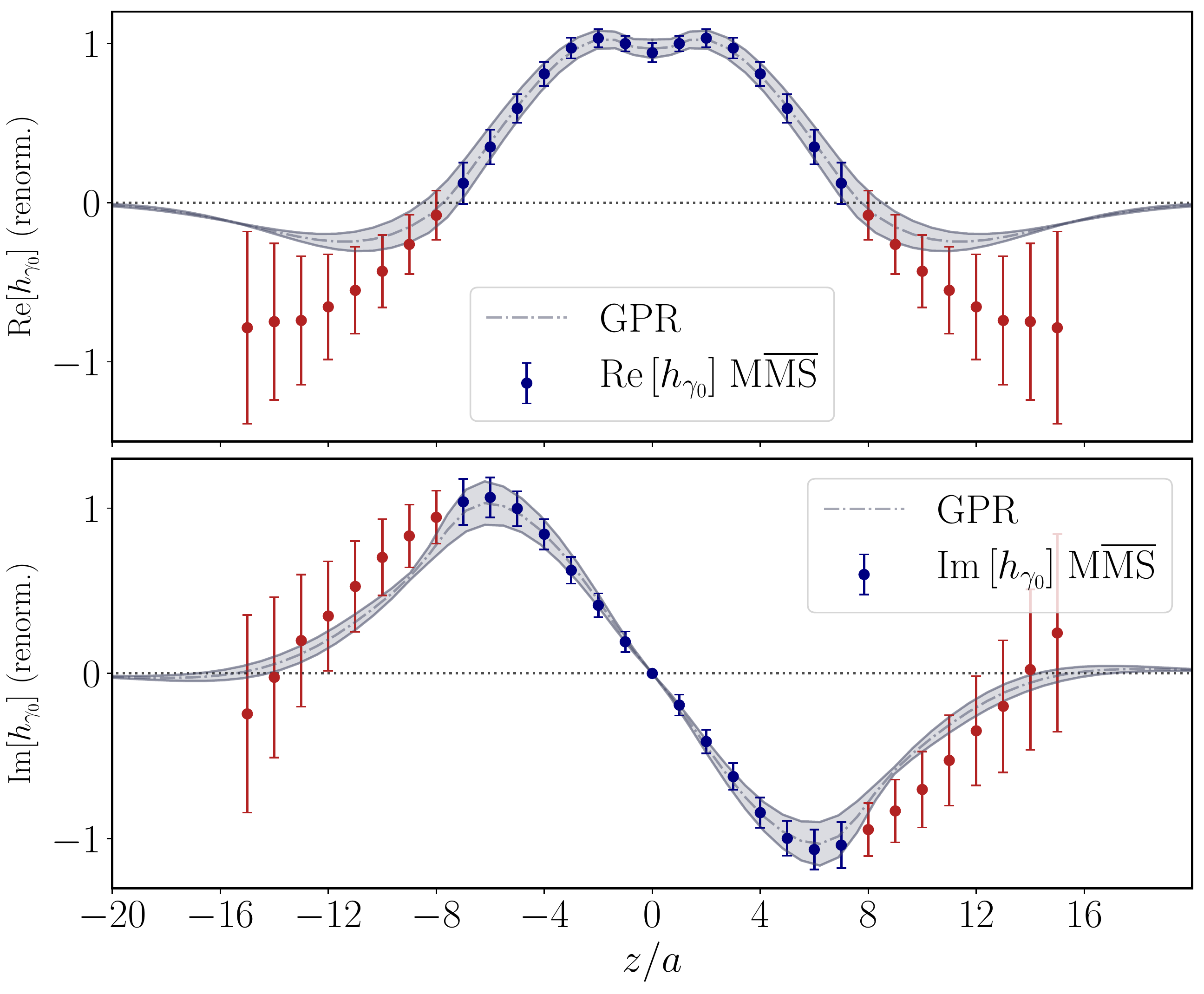}
    \caption{Lattice data for the matrix element of the unpolarized operator together with the non parametric regression performed with nonzero prior. Only points shown with  blue circles are in the fit.}
    \label{fig:GPR_P}
\end{figure}

As previously stated, we use as an alternative prior mean the function of Eq.~\eqref{eq:FT_shift_gauss} with $\mu=0.315$, $\sigma=0.23$ and $c=2.22$ in order to cross-check our results. The outcome of the nonparametric regression with nonzero mean prior is shown  in Fig.~\ref{fig:GPR_P}, while the lower panel of Fig.~\ref{fig:qPDF_comparison} shows the FT of this function together with the resulting BGFT. As pointed out in Sec.~\ref{sec:prior_choice}, the choice of the prior mean function modifies the result of the GPR in the region where there is a low density of data points. In this specific case, it slightly modifies the decay rate in the large-$z$ region, further reducing the amplitude of the remnant oscillations present in the BGFT. However, the effect of the prior mean is not observable in the light-cone PDF that is still compatible with the reconstruction obtained with iDFT, as shown in the lower panel of Fig. \ref{fig:matched_comparison}. \\ The  choice of the prior  within the BGFT approach as well as the choice of the cutoff $z_{max}$ introduce systematic uncertainties.  In this work, we show that the differences between the light-cone PDFs obtained with two different priors $\mu_P(z)$ are negligible, suggesting that the systematic effect due to the choice of the model for the matrix elements outside the fit region does not lead to large effects.  

\section{Conclusions}\label{sec:conclusions}
In this work we address   the nonphysical oscillations, which appear in the computation of PDFs from lattice QCD simulations.
Due to the lattice discretization, the continuous Fourier transform 
in Eq.~\eqref{eq:quasi-pdf} 
 cannot be computed. Moreover, the continuous FT cannot  be replaced by the discrete Fourier transform, since this would require periodicity of the matrix element.  
 To obtain a continuous reconstruction of the FT, 
 we can employ the analytical continuation of the DFT, defined for continuous real values of $x$. However, this transform,  would require the knowledge of the discretized matrix element for arbitrary large values of the Wilson line length $|z|$. The commonly adopted solution in the literature is to assume that the matrix element goes to zero for $|z|>z_{\rm max}$. This assumption leads to the definition of the interpolated DFT (iDFT) that, in contrast to the DFT, provides a continuous frequency domain function that is suitable for computing the light-cone PDF. However, the iDFT is only an interpolation of the DFT, consisting of a linear combination of 
Dirichlet kernels. As a consequence, despite being a continuous function, 
it is still afflicted by the same problems that appear in  DFT (aliasing and leakage) that hinder the evaluation of the transform. These considerations are in accordance with the results 
obtained in a recent paper~\cite{Karpie:2019eiq}, and suggest that the nonphysical oscillations observed in the PDFs computed from lattice QCD 
matrix element may be due to the discretization of the Fourier Transform. 
The problems afflicting the iDFT can be solved by reconstructing a continuous form of the renormalized matrix element defined over the whole domain before evaluating the continuous FT. 
In order to obtain such a continuous reconstruction we employ 
Gaussian process regression.
It consists of a Bayesian nonparametric 
regression that leverages on the smoothness properties of the renormalized matrix element function, and its asymptotic behavior.
Another property that makes GPR a useful tool for quasi-PDF 
computations is that the Fourier transform of the regression result is analytically computable.

We demonstrate the applicability of this approach in Sec. \ref{subsec:mock_test} using  a mock data set 
generated from a function whose FT is known in closed form. Even though this mock data set 
possesses the same limitations characterizing lattice QCD data, the 
Bayes-Gauss-Fourier Transform  is able to capture all the relevant feature of the analytical FT.

The method is applied to the $\mbox{M}\overline{\mbox{MS}}$ renormalized matrix 
element of the unpolarized PDF, computed on the {\it cA2.09.48} ensemble with $N_f=2$ flavors 
of quarks, lattice size $48^3\times 96$ and a source-sink time separation 
$t_s=12a\approx 1.1\mbox{ fm}$. The BGFT shows a significant reduction of 
nonphysical oscillations in the large $\abs{x}$ region, while it is compatible 
within error bars with the DFT transform for $x\in[-1,1]$. However, the improved 
behavior of the quasi-PDF is small for the physical PDF, where no
substantial deviation can be detected as compared to the PDF obtained with the iDFT.

This finding 
suggests that the presence of the nonphysical 
negative values in the light-cone PDF for $x<0$ cannot be ascribed  to the discrete Fourier transform, or at least, this cannot be the sole 
cause of this behavior. 
Thus, it seems to become mandatory to either reach higher nucleon boosts 
in lattice calculations or compute higher perturbative orders for the 
matching and the conversion between renormalization schemes. Also, 
higher twist effects need to be understood. 
Nevertheless, in this paper we have presented an alternative way to analyze the renormalized 
matrix element, which lead to quasi-PDFs with suppressed oscillation. This approach  can thus provide 
a valuable cross-check of lattice computations of parton distribution functions.

\section*{Acknowledgements} \label{sec:acknowledgements}
We would like to thank all the members of the Extended Twisted Mass Collaboration for providing a stimulating forum for cooperation. We thank J. Green for valuable input.  This project has received funding from the Marie Sk\l{}odowska-Curie   European Joint Doctorate  program STIMULATE of the European Commission under grant agreement No 765048. G. I and F. M. are funded under this program.
\appendix
\section{Matching kernels} \label{app:matching_procedure}
In this work we employ the matching formula
\begin{align}
  q(x,\mu)=&\int_{-\infty}^{\infty}\frac{d \xi}{\abs{\xi}}C\left(\xi,\frac{\xi \mu }{xP_3}\right)\tilde{q}\left( {x\over \xi} ,\mu,P_3 \right)\\
  &\int_{-\infty}^{\infty}\frac{dy}{\abs{y}}C\left(\frac{x}{y},\frac{\mu}{y P_3}\right)\tilde{q}\left( y,\mu,P_3 \right),\nonumber
\end{align}
where the matching kernel $C\left(\xi,\eta\right)$ is computed in the $\MMSb$ scheme to one-loop in perturbation theory~\cite{Alexandrou:2019lfo}. At  leading order, the quasi-PDF is equivalent to the light-cone PDF, while the next-to-leading order (NLO) term $\mathcal{O}(\alpha_s)$ is nontrivial. To NLO the matching is given by
\begin{equation}
  C\left(\xi,\eta\right)=\delta \left(1-\xi \right)+C^{\NLO}(\xi,\eta)_{+(1)}.
\end{equation}

The subscript $+(1)$ denotes the plus-prescription at $\xi=1$, which implies
\begin{equation}
\begin{split}
  &q(x,\mu) = \tilde{q}(x,\mu,P_3)+\\
  &+\int_{-\infty}^{\infty}\frac{d \xi}{\abs{\xi}}C^{\NLO}\left(\xi,\frac{\mu \xi}{x P_3}\right)\tilde{q}\left( {x\over \xi} ,\mu,P_3 \right)\\
  &-\tilde{q}\left( x ,\mu,P_3 \right) \int_{-\infty}^{\infty}\frac{d \xi}{\abs{\xi}}C^{\NLO}\left(\xi,\frac{\mu }{x P_3}\right).
\end{split}
\end{equation}

\begin{widetext}
The next-to-leading order term of the matching kernel for the unpolarized case reads
  \begin{equation}\label{quasi_matching_MMSbar}
 \begin{split}
     &C^{\NLO}\left(\xi, \eta \right)_{+(1)}=  \\
&+{\alpha_sC_F\over 2\pi}
\begin{cases}
\displaystyle \left({1+\xi^2\over 1-\xi}\ln \left({\xi\over \xi-1}\right) + 1 + {3\over 2\xi}\right)_{+(1)}, &\, \xi>1,
\nonumber \\[10pt]
\displaystyle \left({1+\xi^2\over 1-\xi}\ln\left[{1\over \eta^2}\big(4\xi(1-\xi)\big)\right] - {\xi(1+\xi)\over 1-\xi} \right)_{+(1)},&\, 0<\xi<1,
\nonumber \\[10pt]
\displaystyle  \left(-{1+\xi^2\over 1-\xi}\ln \left({-\xi\over 1-\xi}\right) - 1 + {3\over 2(1-\xi)}\right)_{+(1)}, \quad
&\, \xi<0.
\end{cases}
 \end{split}  
  \end{equation}
\end{widetext}

Moreover, in Sec. \ref{sec:phenomenological_data}  we apply the \emph{inverse} matching procedure to obtain the matrix elements starting from the phenomenological PDFs. Given the light-cone PDF, the quasi-PDF operator can be computed as
\begin{equation}\label{eq:inverse_matching}
  \tilde{q}\left( y,\mu,P_3 \right)=\int_{-1}^{1}\frac{dy}{\abs{y}}\tilde{C}\left(\frac{x}{y},\frac{\mu}{y P_3}\right)q\left( y,\mu\right). 
\end{equation}
The two kernels $C(\xi,\eta)$ and $\tilde{C}(\xi,\eta)$ only differ in the overall sign of the next-to-leading order term, which is minus for the inverse matching. A detailed description of the inverse matching procedure and the technical aspects of the computation of the integral in Eq.~\eqref{eq:inverse_matching} we refer to Ref.~\cite{Cichy:2019ebf}.

\bibliographystyle{apsrev4-1}
\bibliography{bibliography}

\end{document}